\documentclass[
superscriptaddress,
pra,
twocolumn,
showpacs,
showkeys,
aps,
floatfix,
]{revtex4-1}

\usepackage{amsmath}
\usepackage{graphicx}
\usepackage{bm}
\usepackage{color}

\newcommand{\rmc}{\mathrm{c}}
\newcommand{\rmd}{\mathrm{d}}
\newcommand{\rme}{\mathrm{e}}
\newcommand{\rmi}{\mathrm{i}}

\newcommand{\tr}{{\mathrm tr}}
\newcommand{\tre}{\tr_\rme}

\newcommand{\One}{\openone}

\newcommand{\la}{\langle}
\newcommand{\ra}{\rangle}
\newcommand{\lla}{\left\langle}
\newcommand{\rra}{\right\rangle}

\usepackage{hyperref}

\begin{document}
\title{Single qubit decoherence under a separable coupling to a random matrix 
  environment}

\author{M. Carrera}
\affiliation{Facultad de Ciencias, Universidad Aut\' onoma del Estado de
  Morelos, Avenida Universidad 1001, C.P. 62209, Cuernavaca, Morelos, M\' exico.}
\affiliation{Instituto de Ciencias F\'\i sicas, Universidad Nacional
  Aut\' onoma de M\' exico, Avenida Universidad s/n, 62210 Cuernavaca, Morelos,
  M\' exico.}

\author{T. Gorin}
\affiliation{Departamento de F\'\i sica, Universidad de Guadalajara,
  Blvd. Marcelino Garc\'\i a Barragan y Calzada Ol\'\i mpica, 
  C.P. 44840, Guadalajara, Jal\'\i sco, M\' exico.}

\author{T. H. Seligman}
\affiliation{Instituto de Ciencias F\'\i sicas, Universidad Nacional
  Aut\' onoma de M\' exico, Avenida Universidad s/n, 62210 Cuernavaca, Morelos,
  M\' exico.}
\affiliation{Centro Internacional de Ciencias A.C., Avenida Universidad s/n, 62131 Cuernavaca, Morelos, M\' exico.}

\begin{abstract}
This paper describes the dynamics of a quantum two-level system (qubit) 
under the influence of an environment modeled by an ensemble of random 
matrices. In distinction to earlier work, we consider here separable couplings
and focus on a regime where the decoherence time is of the same order of 
magnitude than the environmental Heisenberg time. We derive an analytical 
expression in the linear response approximation, and study its accuracy by 
comparison with numerical simulations. We discuss a series of unusual 
properties, such as purity oscillations, strong signatures of spectral 
correlations (in the environment Hamiltonian), memory effects and symmetry 
breaking equilibrium states. 
\end{abstract}


\maketitle

\section{\label{I} Introduction}

The general idea of random matrix environments consists in formulating the 
dynamics of an open quantum system (henceforth called the ``central system'') 
in terms of the reduced dynamics of a Hamiltonian system which consists of 
central system and environment.
The intention is to use techniques from random matrix theory (RMT) to integrate
out the dynamics in the environment, such that from a computational point of 
view, one only needs to deal with the degrees of freedom of the central system.
Since their introduction in Ref.~\cite{LutWei99}, such RMT formulations have 
received more and more attention. As a result, a variety of different models 
have been proposed. Some studies have been concentrating on the strong coupling 
regime~\cite{GS02b,GS03,Zni11}. 
A generic RMT coupling with finite coupling strength was discussed 
in~\cite{PGS07}. In all these cases decoherence has been discussed principally
in terms of the mean purity, averaged over the RMT ensemble. Instead,
in~\cite{GPKS08} and~\cite{VinZni12} the average density matrix was 
calculated. Somewhat different models based on random matrices 
are given in~\cite{NoAlJe09,NoAlJe11,Bruzda2009320,PhysRevA.87.022111,Wang:12} 

In the present paper we wish to present a more detailed discussion, expanding
the coupling into separable terms, of which, up to now only the dephasing term
has been considered explicitly.
Only in this case, the problem can be reduced to a fidelity problem in 
the near environment~\cite{GCZ97,GPSS04}. Our RMT model may be 
compared to the master equation for resonance fluorescence~\cite{KimMan76}, or 
to the Jaynes-Cummings model~\cite{jaynescummings1963}, where a two-level 
system is coupled to a single harmonic oscillator mode. When compared to the 
first, the important difference lies in the fact that we have a finite level 
spacing and assume the decoherence time to be of the order of the environmental 
Heisenberg time. When compared to the second, the difference is, that we assume 
a complicated many-body environment about which we know little, essentially 
with properties similar to those assumed for the nucleus by 
Wigner~\cite{Wigner1967}. In this spirit the random matrix environment can be 
considered as the representation of maximal ignorance about this 
environment~\cite{Balian}.
Alternatively, we may consider the RMT environment as a model for a chaotic 
system as considered in~\cite{Casati,BGS84}, which happens to serve as the 
environment for a two level system. Interestingly, 
Ref.~\cite{Lem12} reports the experimental realization of a possibly suitable 
quantum chaotic system.

Our model with separable but otherwise random coupling shows strikingly 
different and unusual phenomena, depending on the term we use: (i) It can show 
a strong sensitivity to spectral correlations, which according to the quantum 
chaos hypothesis~\cite{BerTab77,Casati,BGS84} may allow to distinguish chaotic 
and integrable environments, on the basis of the decoherence process. (ii) In 
spite of the fact that our model defines a unital quantum map~\cite{BenZyc06} 
for the qubit, it may lead to purity oscillations. (iii) Our model 
shows an intriguing case of symmetry breaking by the stationary states, as well
as memory effects. Some of these properties may be related to different 
signatures of quantum non-Markovianity, introduced 
recently~\cite{BrLaPi09,RiHuPl10,Zni11}.

The paper is organized as follows: In Sec.~\ref{G} we introduce our model of an
RMT environment with separable coupling, apply the linear response 
approximation and perform the averages over the random matrices. Sec.~\ref{S} 
specializes the model to the case of a single qubit as the central 
system. In Sec.~\ref{N} we present our numerical simulations, the comparison 
with the linear response approximation, and discuss the very particular 
features for different separable couplings. In Sec.~\ref{C} we present our
conclusions. Various integrals related to the ensemble averages within the 
linear response approximation are calculated in the appendix.

\section{\label{G} General model}

The full Hilbert space $\mathcal{H}_\rmc \otimes \mathcal{H}_\rme$ is divided 
into the Hilbert space of the central system $\mathcal{H}_\rmc$ and that of the
environment $\mathcal{H}_\rme$. We assume that the dynamics in the whole 
Hilbert space is unitary, governed by the Hamiltonian
\begin{equation}
H_\lambda = H_0 + \lambda V \; , \qquad H_0 =  H_\rmc \otimes \One 
   + \One \otimes  H_\rme \; ,
\label{G:Hlam}\end{equation}
where the real and non negative parameter $\lambda$ denotes the strength of the
coupling between the central system $H_\rmc$ and the environment $H_\rme$. For
definiteness, we assume $H_0$ and $\lambda$ to have units of energy, while the
perturbation $V$ itself is dimensionless. We measure time $t_{\rm ph}$ in units 
of $\hbar/d_0 = 2\pi\, t_{\rm H}$ where $t_{\rm H}$ is the Heisenberg time of 
the environment and $d_0$ is the average level spacing in the spectrum of 
$H_\rme$. In terms of the dimensionless time $t= t_{\rm ph} d_0/\hbar$, the 
evolution operator may be written as $U(t)= \exp[-\rmi H_\lambda t/d_0]$. In 
order to complete the change to dimensionless quantities, we choose
$\lambda = d_0\, \mu$, $H_\rmc = d_0\, h_\rmc$ and $H_\rme = d_0\, h_\rme$
such that
\begin{equation}
  H_\lambda = d_0\, h_\mu \; , \qquad h_\mu= h_0 + \mu\; V \; .
\label{G:HamRescaling}\end{equation}
Hence, $U(t)= \exp[-\rmi h_\mu\, t]$, and 
$h_0= h_\rmc\otimes\One + \One\otimes h_\rme$. In contrast to earlier 
work~\cite{PGS07,GPKS08}, the coupling operator $V$ is now assumed to be 
separable:
\begin{equation}
V= v_\rmc \otimes V_\rme \; ,
\end{equation}
where $V_\rme$ is a random matrix chosen from one of the Gaussian invariant
ensembles~\cite{Meh2004}. In what follows, we may want to distinguish the case where $v_\rmc$
commutes with $h_\rmc$ (dephasing coupling~\cite{GCZ97,GPSS04}) and the case 
where it does not commute.

At a first stage, we follow~\cite{GPKS08} in order to calculate the density 
matrix of the central system
\begin{equation}
\varrho_\rmc(t) = \tre\left[ \rme^{-\rmi h_\mu t}\, \varrho_0\,
   \rme^{\rmi h_\mu t}\right]\; , \quad 
   \varrho_0 = \varrho_\rmc \otimes \varrho_\rme \; .
\label{G:reducedDyn}\end{equation}
This yields still without any approximation (for details see Sec.~2 
of~\cite{GPKS08})
\begin{equation}
\varrho_\rmc(t) = u_\rmc\; \tilde\varrho_\rmc(t)\; u_\rmc^\dagger\; , 
\qquad \tilde\varrho_\rmc(t)= \tre\big [\, \varrho_M(t)\, \big ] \; ,
\label{trhoc}\end{equation}
where $\varrho_M(t)$ stands for the density operator in the interaction 
picture:
\begin{equation}
\varrho_M(t)= M_\mu(t)\; \varrho_0\; M_\mu(t)^\dagger \; , \quad
M_\mu(t)= \rme^{\rmi h_0\, t}\; \rme^{-\rmi h_\mu\, t} \; .
\label{eq:rhoecho}\end{equation}
In the following sections, we introduce the linear response approximation,
where we will be able to perform the averages over the random matrices in 
$\varrho_M(t)$.

\subsection{\label{GL} Linear response approximation}

In order to apply linear response theory, we  develop $M_\mu(t)$ into 
its Dyson series, and consider terms up to second order in $\mu$. This yields
\begin{equation}
\tilde\varrho_\rmc(t) = \varrho_\rmc - \mu^2\, \tre\big [\, 
   J(t)\, \varrho_0 + \varrho_0\, J(t)^\dagger + I(t)\, \varrho_0\, I(t)\, 
   \big ] \; ,
\label{GL:rhoc}\end{equation}
where
\begin{equation}
I(t)= \int_0^t\rmd s\; \tilde V(s) \; , \quad
J(t)= \int_0^t\rmd s\int_0^s\rmd s'\; \tilde V(s)\, \tilde V(s') \; .
\end{equation}
Due to the separability of $V$, we find for its representation in the 
interaction picture
\begin{align}
\tilde V(t) &= \rme^{\rmi\, h_\rmc\, t}\otimes \rme^{\rmi\, h_\rme\, t}\; 
   v_\rmc\otimes V_\rme\; 
   \rme^{-\rmi\, h_\rmc\, t}\otimes \rme^{-\rmi\, h_\rme\, t}\notag\\
 &= u_\rmc^\dagger\; v_\rmc\; u_\rmc\otimes u_\rme^\dagger\; V_\rme\; 
     u_\rme =\tilde v_c(t)\; \otimes \; \tilde V_\rme(t) \; .
\end{align}

\subsection{\label{GR} Random matrix averages}

Our model contains two random matrices, the coupling matrix $V_\rme$ (with
matrix elements $V_{jl}$) and the environment Hamiltonian $h_\rme$. In this
section we consider the case where $V_\rme$ is chosen from the Gaussian unitary
ensemble (GUE) and the Gaussian orthogonal ensemble (GOE), while for $h_\rme$, 
we only assume that it can be diagonalized leaving the respective ensemble 
invariant. Under such conditions we can arrive at a very compact expression
for the average reduced state $\la\varrho_\rmc(t)\ra$. Note however that
later on we will restrict ourselves to the GUE.

To obtain the ensemble average of the reduced state $\la\varrho_\rmc(t)\ra$, we 
first perform the average over $V_\rme$ and later that over $h_\rme$. In view 
of Eq.~(\ref{GL:rhoc}), we divide that calculation in two parts, the 
calculation of $\la J(t)\ra$ and $\la I(t)\, \varrho_0\, I(t)\ra$, 
respectively.
\begin{align}
&\la J(t)\ra =
  \int_0^t\rmd s\int_0^s\rmd s'\; \tilde v_c(s) \tilde v_c(s') \notag\\
&\qquad\otimes 
  \lla \sum_{jln} |j\ra\; \rme^{\rmi (E_j - E_l) s}\; V_{jl}\; 
  \rme^{\rmi (E_l - E_n) s'}\; V_{ln}\; \la n| \rra \notag\\
&\quad= \int_0^t\rmd s\int_0^s\rmd s'\; \tilde v_c(s)\; \tilde v_c(s') \notag\\
&\qquad\otimes 
  \sum_{jln} |j\ra\; \rme^{\rmi (E_j - E_l) s}\; 
  \delta_{jn}\; \rme^{\rmi (E_l - E_n) s'}\; \la n| \; .
\end{align}
The term $\delta_{jl}\delta_{ln} = \delta_{jn}\delta_{ln}$ is present in the 
GOE case ($\beta = 1$), but absent in the GUE case ($\beta = 2$). Therefore,
\begin{align} 
\la J(t)\ra &= \int_0^t\rmd s\int_0^s\rmd s'\; \tilde v_c(s) \tilde v_c(s') 
   \notag\\
&\qquad\otimes \sum_n |n\ra\la n| \left( \textstyle
  3 - \beta + \sum_{l\ne n} \rme^{\rmi (E_n - E_l) (s-s')}\right) 
\notag\\
&= \int_0^t\rmd s\int_0^s\rmd s'\; c(s - s')\; \tilde v_c(s) \tilde v_c(s') 
   \otimes \One\; ,
\end{align}
where $\beta$ is the so called Dyson parameter, and
\begin{equation}
c(t)= 3 - \beta + \delta\left(\frac{t}{2\pi}\right) 
   - b_2\left(\frac{t}{2\pi}\right)\; .
\label{correlfun}\end{equation}
The calculation presented here is completely analogous to the one in 
Ref.~\cite{GPS04}, where we worked out the linear response result for fidelity 
decay in a similar RMT model. The function $b_2(t)$ is the so called 
two-point form factor~\cite{Meh2004}. For the average over the second term we 
find
\begin{align}
&\la I(t)\, \varrho_0\, I(t)\ra 
  = \int_0^t\rmd s\int_0^t\rmd s'\; \tilde v(s)\; \varrho_\rmc\; \tilde v(s')
 \sum_{jln} \rme^{\rmi E_j (s - s')}\;\qquad \notag\\
&\qquad\times\; (\delta_{ln} + \delta_{jn}\delta_{lj})\;
   \rme^{-\rmi (E_l s - E_n s')} \; \varrho^\rme_{ln} \; ,
\end{align}
where we have used that $\varrho_0$ is separable: 
$\varrho_0 = \varrho_\rmc \otimes \varrho_\rme$. Since 
$\delta_{ln} + \delta_{jn}\delta_{lj} = \delta_{ln} (1 + \delta_{jn})$, we
obtain
\begin{align}
&\la I(t)\, \varrho_0\, I(t)\ra= \int_0^t\rmd s\int_0^t\rmd s'\; \tilde v(s)\; 
   \varrho_\rmc\; \tilde v(s')\notag\\
&\qquad\qquad\qquad\times\; \sum_{jn} \rme^{\rmi (E_j-E_n) (s - s')}\;
   (1 + \delta_{jn})\; \varrho^\rme_{nn}
\end{align}
in the GOE case, and
\begin{align}
&\la I(t)\, \varrho_0\, I(t)\ra = \int_0^t\rmd s\int_0^t\rmd s'\; 
   \tilde v(s)\; \varrho_\rmc\; \tilde v(s') \notag\\
&\qquad\qquad\quad\times\; \left[ 3 - \beta + \textstyle
   \sum_n \varrho^\rme_{nn}\sum_{j\ne n}
   \rme^{\rmi (E_j-E_n) (s - s')}\right] \notag\\
&\qquad\qquad = \int_0^t\rmd s\int_0^t\rmd s'\; c(s - s')\; 
   \tilde v(s)\; \varrho_\rmc\; \tilde v(s')\; .
\end{align}
in general. With both results, we obtain:
\begin{align}
&\tilde\varrho_\rmc(t) = \varrho_\rmc \label{eq-principal}\\
&\quad  - \mu^2\; \int_0^t\rmd s\int_0^s\rmd s'\; c(s-s')\; 
   \big [\, \tilde v_c(s)\, ,\,
      [ \tilde v_c(s') , \varrho_c ]\, \big ] \; .
\notag\end{align}

\section{\label{S} Single qubit central system}

For a general two level system as central system, we may assume without 
restriction that $h_\rmc$ is given by
\begin{equation}
h_\rmc = \frac{\Delta}{2}\; \sigma_z \; ,
\end{equation}
with $\Delta$ being the energy difference between the two eigen states of the 
qubit, measured in units of the mean level spacing $d_0$ in the environment;
cf. Eq.~(\ref{G:HamRescaling}). The matrix $\sigma_z$ is one of the three Pauli
matrices~\cite{BenZyc06}
\begin{alignat}{2}
\sigma_x &= \begin{pmatrix} 0 & 1\\ 1 & 0\end{pmatrix} \; , \quad
\sigma_y &= \begin{pmatrix} 0 & -\rmi\\ \rmi & 0\end{pmatrix} \; , \quad
\sigma_z &= \begin{pmatrix} 1 & 0\\ 0 & -1\end{pmatrix} \; .
\label{PauliMats}\end{alignat}
We consider the single qubit coupled to a large environment via a 
separable coupling, where the qubit-part of the coupling either commutes with 
the system part ($v_\rmc = \sigma_z$) or not. The former case is called 
``dephasing coupling''~\cite{GCZ97,GPSS04}, while in the latter case, we 
choose $v_\rmc= \sigma_x$ (this is equivalent to any linear combination of 
$\sigma_x$ and $\sigma_y$). Note that we can obtain the general RMT Hamiltonian 
for the central system as a random linear combination of these terms.

\subsection{\label{PZ} Dephasing coupling, \boldmath $v_\rmc = \sigma_z$} 

In this case, the evolution in the environment only depends parametrically on 
the state of the central system (dephasing situation)~\cite{GCZ97,GPSS04}. That
allows one to find an exact analytical expression for the RMT average of
the reduced dynamics in terms of fidelity decay~\cite{StoSch04b,StoSch05}. In
this section, we nevertheless use the linear response approximation. This 
allows us to make contact with fidelity calculation in Ref.~\cite{GPS04}, and
it serves us as a test case, before embarking on the case $v_\rmc= \sigma_x$,
which is at the center of our interest.

For $v_\rmc = \sigma_z$, we find for the coupling operator in the interaction
picture:
\begin{equation}
v_\rmc = \left(\begin{array}{cc} 1 & 0\\ 0 & -1\end{array}\right) \; , \quad
\tilde v_\rmc(s) = \tilde v_\rmc(s') 
  = \left(\begin{array}{cc} 1 & 0\\ 0 & -1\end{array}\right) \; .
\end{equation}
In order to evaluate the linear response expression in 
Eq.~(\ref{eq-principal}), we need the commutators
\begin{align}
[\tilde v_\rmc(s')\, ,\, \varrho_\rmc] &= 
   \left(\begin{array}{cc} 1 & 0\\ 0 & -1\end{array}\right)
   \left(\begin{array}{cc} a & b^*\\ b & c\end{array}\right)
 - \left(\begin{array}{cc} a & b^*\\ b & c\end{array}\right)
   \left(\begin{array}{cc} 1 & 0\\ 0 & -1\end{array}\right) \nonumber\\
 &= \left(\begin{array}{cc} 0 & 2b^*\\ -2b & 0\end{array}\right) \; ,
\end{align}
and
\begin{align}
&[\tilde v_\rmc(s)\, ,\, \ldots \, ] = 
   \left(\begin{array}{cc} 1 & 0\\ 0 & -1\end{array}\right)
   \left(\begin{array}{cc} 0 & 2b^*\\ -2b & 0\end{array}\right)\notag\\
&\qquad\quad - \left(\begin{array}{cc} 0 & 2b^*\\ -2b & 0\end{array}\right)
   \left(\begin{array}{cc} 1 & 0\\ 0 & -1\end{array}\right)
= \left(\begin{array}{cc} 0 & 4b^*\\ 4b & 0\end{array}\right) \; ,
\end{align}
where we introduced the notation ``$\ldots$'' on the left side to denote 
$[\tilde v_\rmc(s')\, ,\, \varrho_\rmc]$. As expected in this ``dephasing''
case, only the off-diagonal elements of the central system's density matrix are 
affected. 

If the initial state is an eigenstate of $\sigma_z$, then there is no evolution 
at all. On the other hand, for a pure state in the $(x,y)$-plane of 
the Bloch sphere (for the sake of definiteness, we choose the symmetric 
eigenstate of $\sigma_x$), we obtain
\begin{equation} 
|\psi\ra = \frac{1}{\sqrt{2}}\; \big (\, |0\ra + |1\ra\, \big ) \; , \quad
\varrho_\rmc = |\psi\ra\la\psi| = \frac{1}{2}\begin{pmatrix} 1& 1\\ 1& 1
   \end{pmatrix} \; ,
\end{equation}
such that $a=b=c= 1/2$. Therefore:
\begin{align}
\la\tilde\varrho_\rmc^{21}(t)\ra &= \frac{1}{2} - \mu^2
   \int_0^t\rmd s\int_0^s\rmd s'\; c(s - s')\; 2\notag\\
 &= \frac{1}{2}\left[ 1 - 4\mu^2\; C_{\rm fid}(t)\right] \; .
\label{S:defCfid} 
\end{align}
The function $C_{\rm fid}(t)$ is evaluated in App.~\ref{apED} with the result:
\begin{equation}
C_{\rm fid}(t)
=\begin{cases} \pi t + t^3/(12\pi) &: t < 2\pi\; ,\\
      t^2/2 + 2\pi^2/3 &: t > 2\pi\; ,
    \end{cases} 
\end{equation}
in agreement with the general result in Ref.~\cite{GPS04}.

\subsection{\label{PX} Non-commuting coupling, \boldmath $v_\rmc = \sigma_x$}

In general, the coupling operator $v_\rmc$ may be written as a linear 
combination of the Pauli matrices, defined in Eq.~(\ref{PauliMats}). Since we 
have analyzed the case, where $v_\rmc$ commutes with the system Hamiltonian, we 
are now looking for cases where $v_\rmc$ does not commute. In fact, we would 
like to avoid that the coupling operators has any commuting component. This 
restricts $v_\rmc$ to a linear combination of $\sigma_x$ and $\sigma_y$. Since
the only other non-trivial central system operator in the Hamiltonian is 
$\sigma_z$, we may choose $v_\rmc = \sigma_x$ without restricting generality.
Hence, the dynamics in our model only depends on the initial state 
$\varrho_\rmc$ of the qubit, i.e. its orientation in the Bloch sphere, with 
respect to the system Hamiltonian $\sigma_z$ and the coupling operator 
$\sigma_x$. The linear response calculation then gives:
\begin{equation}
\tilde v_\rmc(s') = \begin{pmatrix} \rme^{\rmi\Delta s'/2} & 0 \\
                         0 & \rme^{-\rmi\Delta s'/2} \end{pmatrix} 
\begin{pmatrix} 0 & 1\\ 1 & 0\end{pmatrix}
  \begin{pmatrix} \rme^{-\rmi\Delta s'/2} & 0 \\
                          0 & \rme^{\rmi\Delta s'/2} \end{pmatrix} \; ,
\end{equation}
which results in
\begin{equation}
\tilde v_\rmc(s') = \begin{pmatrix} 0 & \kappa_1\\ \kappa_1^* & 0\end{pmatrix}
\; , \quad 
\tilde v_\rmc(s) = \begin{pmatrix} 0 & \kappa_2\\ \kappa_2^* & 0\end{pmatrix}
\; ,
\end{equation}
where $\kappa_1 = \rme^{\rmi\Delta s'}$, and 
$\kappa_2 = \rme^{\rmi\Delta s}$. For the commutators we then obtain:
\begin{align}
[\tilde v_\rmc(s')\, ,\, \varrho_\rmc] &= \begin{pmatrix} 
   \kappa_1 b - \kappa_1^* b^* & -\kappa_1 (a-c)\\
   \kappa_1^* (a-c) & -(\kappa_1 b - \kappa_1^* b^*)\end{pmatrix} \notag\\
 &= \begin{pmatrix} Q & -\kappa_1\, R \\
      \kappa_1^*\, R & -Q \end{pmatrix}\; ,
\end{align}
where $Q= \kappa_1 b - \kappa_1^* b^*$ and $R= a - c$. Finally:
\begin{align}
&[\tilde v_\rmc(s)\, ,\, \ldots\, ] =
   \begin{pmatrix} 0 & \kappa_2\\ \kappa_2^* & 0\end{pmatrix}
   \begin{pmatrix} Q & -\kappa_1\, R \\
            \kappa_1^*\, R & -Q \end{pmatrix}\notag\\
&\qquad\qquad - \begin{pmatrix} Q & -\kappa_1\, R \\
               \kappa_1^*\, R & -Q \end{pmatrix}
   \begin{pmatrix} 0 & \kappa_2\\ \kappa_2^* & 0\end{pmatrix} \notag\\
&\qquad = \begin{pmatrix}
    (\kappa_1 \kappa_2^* + \kappa_1^* \kappa_2)\, R & -2\kappa_2\, Q\\
   2\kappa_2^*\, Q & -(\kappa_1 \kappa_2^* + \kappa_1^* \kappa_2)\, R
                                                       \end{pmatrix} \; .
\label{S:dcomutat}\end{align}
In what follows, we derive explicit expressions for the matrix elements of 
the reduced quantum state, for different initial states, first for eigenstates
of $\sigma_x$ (the coupling operator), then for eigenstates of $\sigma_y$, and
finally for eigenstates of $\sigma_z$ (the central system operator).

\subsubsection{Eigenstates of the coupling operator \boldmath $\sigma_x$}

For definiteness, we consider the initial state $\varrho_\rmc$ to be the 
eigenstate of $\sigma_x$ with eigenvalue $+1$. This implies that $a=b=c=1/2$,
$Q= (\kappa_1 - \kappa_1^*)/2 = \rmi\, \sin(\Delta s')$ and $R=0$. Therefore,
Eq.~(\ref{S:dcomutat}) yields
\begin{equation}
[\tilde v_\rmc(s)\, ,\, \ldots\, ] = 2\rmi\, \sin(\Delta s') 
\begin{pmatrix} 0 & -\rme^{\rmi\Delta\, s}\\
    \rme^{-\rmi\Delta\, s} & 0 \end{pmatrix} \; .
\label{S:sigmax1}\end{equation}
Thus, in the present case, we have again a dephasing situation. While the
diagonal elements of the average density matrix remain constant, the 
off-diagonal element becomes
\begin{align}
\la\tilde\varrho_\rmc^{21}(t)\ra &= \frac{1}{2} - 2\rmi\; \mu^2
   \int_0^t\rmd s\; \rme^{-\rmi\Delta\, s}\int_0^s\rmd s'\; c(s-s')\; 
   \sin(\Delta s') \notag\\
&= \frac{1}{2}\big [\, 1 - 4\, \mu^2\; C_x(t) \, \big ] \; ,
\label{S:sigmax2}\end{align}
with the change of variable $s' \to u=s-s'$ the function $C_x(t)$ becomes
\begin{equation}
C_x(t)= \rmi\int_0^t\rmd u\; c(u) \int_u^t\rmd s\; \rme^{-\rmi\Delta\, s}\;
   \sin\, [\Delta(s-u)] 
\label{S:defCx}\end{equation}
is again evaluated in App.~\ref{apE} with the result given in the 
Eqs.~(\ref{apE:resCx0}) and~(\ref{apE:resBx}).

\subsubsection{Eigenstates of \boldmath $\sigma_y$} 

We choose the eigenstate $|\psi\ra = (|0\ra +\rmi\, |1\ra)/\sqrt{2}$ which has
eigenvalue $+1$. Then
\begin{equation}
\varrho_{\rm c}= \frac{1}{2}\begin{pmatrix} 1 & -\rmi \\ \rmi & 1\end{pmatrix}
\end{equation}
such that $Q= \rmi\, (\kappa_1 + \kappa_1^*)/2$ and $R=0$. Hence
\begin{align}
[\tilde v_\rmc(\sigma)\, ,\, \ldots\, ] &= \left(\begin{array}{cc} 
   0 & -\rmi\, \kappa_2\, (\kappa_1 + \kappa_1^*)\\
   \rmi\, \kappa_2^*\, (\kappa_1 + \kappa_1^*) & 0 \end{array}\right) \notag\\
&= 2\rmi\, \cos(\Delta s') \begin{pmatrix} 0 & -\rme^{\rmi\Delta\, s}\\
    \rme^{-\rmi\Delta\, s} & 0\end{pmatrix} \; .
\label{S:doublecomcos}\end{align}
The result is quite similar to the $\sigma_x$ eigenstate case. The diagonal
elements of the average density matrix remain again constant, while the 
non-diagonal element becomes
\begin{equation}
\la\tilde\varrho_\rmc^{21}(t)\ra= \frac{\rmi}{2}\, \big [\, 1 - 4\mu^2\; C_y(t)
  \, \big ] \; ,
\label{S:tivarhoy}\end{equation}
where
\begin{equation}
C_y(t)= \int_0^t\rmd u\; c(u)\int_u^t\rmd s\; \rme^{-\rmi\Delta\, s}\, 
   \cos[\Delta(s-u)]\; .
\end{equation}
The expression for $C_y(t)$ differs from that for $C_x(t)$ in 
Eq.~(\ref{S:defCx}) only in as much as the sine function in Eq.~(\ref{S:defCx})
is replaced here by the cosine function. The result of the evaluation of the 
integrals is given in Eqs.~(\ref{apE:resCy0}) and~(\ref{apE:resBy}).

\subsubsection{Eigenstates of the system Hamiltonian \boldmath $\sigma_z$.}  

For definiteness, we again choose the eigenstate corresponding to the 
eigenvalue $+1$. Thus, the initial state is given by the density matrix
\begin{equation} 
\varrho_{\rm c} = |0\ra\la 0| \; , \quad a=1\; , \quad b=c=0 \; , \quad
   Q= 0\; , \quad R=1 \; . 
\end{equation}
Therefore, we have
\begin{equation}
[\tilde v_\rmc(s)\, ,\, \ldots\, ] = 2\, \cos\, [\Delta (s -s')]\, \begin{pmatrix}
 1 & 0 \\ 0 & -1\end{pmatrix}\; .
\label{S:XiZcoupl}\end{equation}
This means that the non-diagonal elements of the average density matrix will 
remain zero, while 
\begin{equation}
\la\tilde\varrho_\rmc^{11}(t)\ra = 1 - 2\mu^2\; C_z(t) \; , 
\label{S:tirhoz}\end{equation}
in the same way as above, we do a change of variable $s'\to u=s-s'$ which
yields
\begin{equation}
C_z(t)= \int_0^t\rmd u\; c(u)\; \cos(\Delta\, u)\; (t-u)\; 
\end{equation}
and $\la\tilde\varrho_\rmc^{22}(t)\ra= 1 - \la\tilde\varrho_\rmc^{11}(t)\ra$. 
The result of the evaluation of the integral is given in 
Eqs.~(\ref{apE:resC3x0}) and~(\ref{apE:resB3x}).

\section{\label{N} Numerical simulations}

For the simulation of the reduced dynamics defined in Eq.~(\ref{G:reducedDyn}), 
we need the random matrices, $h_\rme$ and $V_\rme$. The matrix $V_\rme$ is a 
GUE matrix normalized such that the off-diagonal elements have unit variance, 
while $h_\rme$ is a diagonal matrix of eigenvalues of a GUE, unfolded to unit 
mean level spacing across the full spectral range. The initial state is a 
separable pure state, while the environmental part is given by a random state, 
chosen to be invariant under arbitrary unitary 
transformations~\cite{Bro81}. 
From a theoretical point of view this choice is entirely equivalent to the 
initial condition $\varrho_\rme = \One/N_\rme$. However, the evolution of a 
pure state can be obtained from the solution of the corresponding 
Schr\" odinger equation, while a mixed initial state would require to
solve a von Neumann equation. The latter is much more time consuming and 
therefore impractical. Unless stated otherwise, we choose $N_\rme = 200$ for 
the dimension of the environment, and for every simulation we average over 
$n_{\rm run} = 300$ realisations.

We consider the reduced dynamics of the qubit in the so called universal 
regime, where the RMT model is expected to yield similar results as
dynamical models, satisfying the quantum chaos conjecture~\cite{BGS84}. That 
means that the time scales considered are of the order of the Heisenberg time 
(in our units, this is at $t=2\pi$), more precisely we require 
$N_\rme^{-1} \ll t/(2\pi) \ll N_\rme$.
Due to these restrictions, we choose the coupling strength $\mu$ sufficiently 
weak, such that the decoherence time fulfills the above inequalities.
The reduced dynamics of the qubit also depends on the level splitting $\Delta$
(in units of $d_0$). In a standard situation, we expect
$1 \ll \Delta \ll N_\rme$, meaning that from the central system point
of view, the spectrum of the environment is both, dense and infinitely large.
However, we will find interesting effects when $\Delta$ is of the order of one 
or even smaller. 

In the dephasing case ($v_\rmc = \sigma_z$), the behavior of the qubit state
is well understood and its dependence on $\Delta$ becomes trivial (cf. 
Sec.~\ref{PZ}). Therefore, we restrict our numerical analysis to the case 
$v_\rmc = \sigma_x$. In what follows, we will discuss the general features of 
the qubit dynamics (in Sec.~\ref{SG}), the effect of spectral correlations (in
Sec.~\ref{SS}), the accuracy of the linear response approximation (in 
Sec.~\ref{SL}), and finally some non-trivial results for the equilibrium state 
-- i.e. the state reached at $t\to\infty$ (in Sec.~\ref{SE}).

Decoherence is measured in terms of purity,
\begin{equation}
P(t)= {\rm tr}\big [\, \la\varrho_\rmc(t)\ra^2\, \big ] \; ,
\label{S:defPur}\end{equation}
calculated from the average density matrix of the qubit obtained from 
averaging over $n_{\rm run}$ realisations of the reduced dynamics defined in
Eq.~(\ref{G:reducedDyn}).

\subsection{\label{SG} General behavior}

In this section, we fix the coupling strength to $\mu=0.1$ and the level
splitting in the qubit Hamiltonian to $\Delta = 1$. Following the 
classification in echo dynamics~\cite{GPSZ06}, $\mu = 0.1$ puts us well 
into the crossover area between the Fermi golden rule and the perturbative 
regimes. Also $\Delta = 1$ is an intermediate choice. For $\Delta \gg 1$ 
we would expect an exponential behavior, similar to open systems described by
quantum master equations. By contrast, for $\Delta \ll 1$, it is actually 
difficult for the central system to couple to the environment, since the 
level repulsion reduces the probability to find transitions which meet the 
excitation energy $\Delta$ for the qubit.

The following figures show results for the matrix elements of the average
density matrix $\la\varrho_\rmc(t)\ra$ of the qubit, and for the purity $P(t)$, 
as defined in Eq.~(\ref{S:defPur}). As initial states for the qubit, we choose
the symmetric eigenstates of the Pauli matrices, {\it i.e.} those eigenstates
which correspond to the eigenvalue $+1$. Thus, we will show three cases, one
for each Pauli matrix, given in Eq.~(\ref{PauliMats}).

\begin{figure}
\includegraphics[width= 0.67\textwidth]{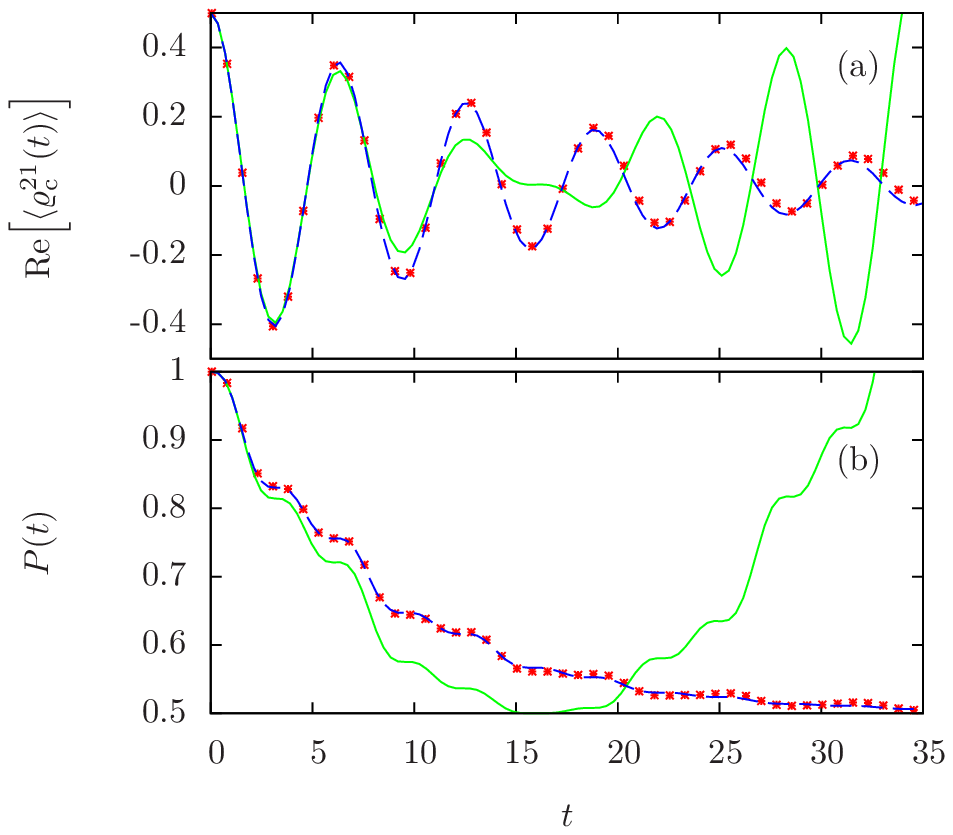}
\caption{(Color online) The reduced dynamics for the symmetric eigenstate of 
$\sigma_x$ as initial state, for $\Delta =1$ and $\lambda = 0.1$ as a function 
of rescaled time $t$, where the Heisenberg time is $t= 2\pi$. Numerical 
simulations (red points) are compared to the linear response theory (green 
solid line) and its exponentiated version (blue dashed line). In panel (a) we 
show the real part of $\la\varrho_\rmc^{21}(t)\ra$, in panel (b) the 
purity of  $\la\varrho_\rmc(t)\ra$.}
\label{S:fig1a}\end{figure}

\paragraph{Eigenstate of \boldmath $\sigma_x$}
In Fig.~\ref{S:fig1a} we show simulations for the reduced dynamics of the qubit
where the initial state is chosen as the symmetric eigenstate of $\sigma_x$.
The results suggest that the evolving qubit state is of the form
\begin{equation}
\la\varrho_\rmc(t)\ra = \frac{1}{2}\begin{pmatrix} 1 & z^*(t)\\ z(t) & 1
   \end{pmatrix} \; ,
\label{S:genrhoc}\end{equation}
which corresponds to a trajectory in the $xy$ plane in the Bloch sphere. For 
this reason, we only consider the non-diagonal element of the qubit state in
panel (a) of Fig.~\ref{S:fig1a}. The deviations of the diagonal elements from
the value $1/2$ are very small and well within the expected statistical error. 
The oscillations which can be seen on panel (a), stem from the central system 
Hamiltonian, which forces the qubit to rotate around the $z$-axis. On panel (b)
we show the purity, which in the present case can be written as
\begin{equation}
P(t)= \frac{1 + |z(t)|^2}{2} \; .
\label{S:purrhoc}\end{equation}
We checked that the somewhat irregular oscillations around an apparently 
exponential decay of $|z(t)|^2$ are not of statistical nature. They reflect the
dynamics of the true RMT average of the reduced dynamics.

For the present case, the linear response approximation yields the expression
given in Eqs.~(\ref{S:sigmax2}) and~(\ref{S:defCx}), plotted as a green solid
line in Fig.~\ref{S:fig1a}. We can see that at least for short times, the 
approximation describes the true evolution very well. For larger times, we see
clear deviations which eventually end up in unphysical behavior. Of course this
has to be expected, since the approximation is based on the truncation of a 
series expansion. Guided by the success of the ``exponentiation'' of the 
linear response approximation in the case of fidelity decay~\cite{GPS04}, we 
try the same trick here, and apply the following replacement to 
Eq.~(\ref{S:sigmax2}):
\begin{equation}
1 - 4\mu^2\; C_x(t) \quad\to\quad \rme^{-4\mu^2\; C_x(t)}\; .
\end{equation}
We refer to this result as the ``exponentiated linear response approximation''
(ELR for short). It is shown by blue dashed lines in Fig.~\ref{S:fig1a}. The 
agreement with the numerical simulations is greatly improved, in the behavior 
of the coherence $\la\varrho_\rmc(t)\ra$, panel (a), as well as the purity 
$P(t)$, panel (b). A more detailed analysis of the accuracy of the ELR is 
provided in Sec.~\ref{SL}.

\begin{figure}
\includegraphics[width= 0.66\textwidth]{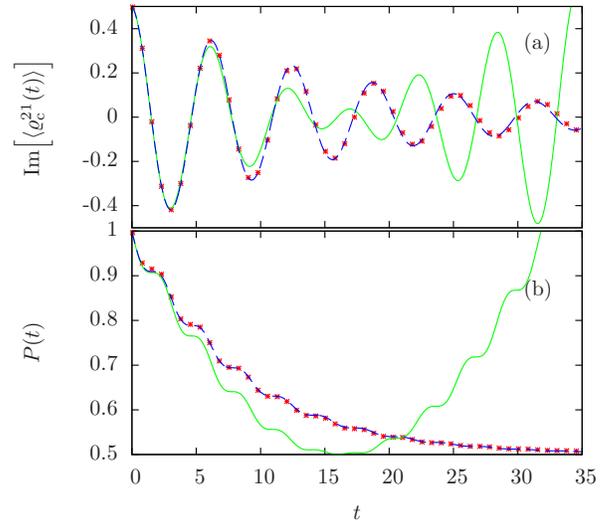}
\caption{(Color online) The reduced dynamics for the symmetric eigenstate of 
$\sigma_y$ as initial state, for $\Delta =1$ and $\lambda = 0.1$ as a function 
of rescaled time, where the Heisenberg time is at $t= 2\pi$. Numerical 
simulations (red points) are compared to the linear response theory (green 
solid line) and its exponentiated version (blue dashed line). Panel (a) shows 
the imaginary part of $\la\varrho_\rmc^{21}(t)\ra$, and (b) the purity of 
$\la\varrho_\rmc(t)\ra$.}
\label{S:fig3-1}\end{figure}

\paragraph{Eigenstate of \boldmath $\sigma_y$}
The simulations shown in Fig.~\ref{S:fig3-1} are entirely analogous to those of
Fig.~\ref{S:fig1a}, except for the different initial state, which is here a
symmetric eigenstate of $\sigma_y$. The behavior of the average reduced density 
matrix is very similar also, and suggests that Eq.~(\ref{S:genrhoc}) 
and~(\ref{S:purrhoc}) apply as well. Note however, that panel (a) of 
Fig.~\ref{S:fig3-1} shows the imaginary part of $\la\varrho_\rmc^{21}(t)\ra$ 
and that the purity in panel (b) shows a slightly different behavior than in
Fig.~\ref{S:fig1a}, as can be seen most clearly at very small times. The 
accuracy of the linear response approximation and its exponentiation is again 
very similar to the previous case. For the exponentiation we apply the 
replacement $1 - 4\mu^2\, C_y(t) \to \exp[-4\mu^2\, C_y(t)]$ to 
Eq.~(\ref{S:tivarhoy}). As can be seen in Fig.~\ref{S:fig3-1}, the agreement 
between numerical simulation (red points) and the ELR result (blue dashed line) 
is surprisingly good.

\begin{figure}
\includegraphics[width= 0.67\textwidth]{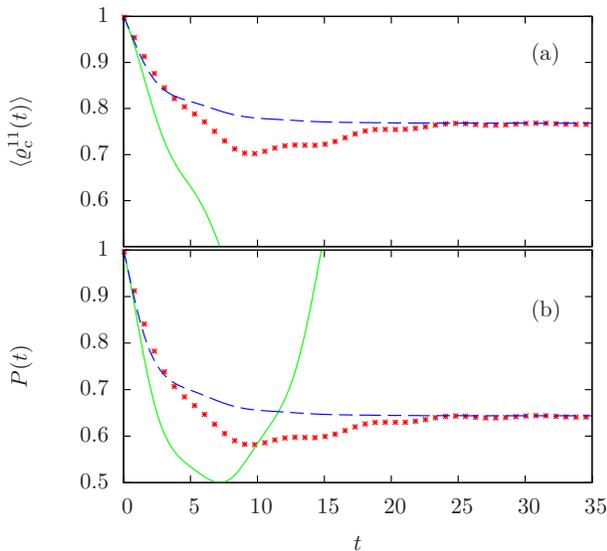}
\caption{(Color online) The reduced dynamics of the qubit for the symmetric 
eigenstate of $\sigma_z$ as initial state, for $\Delta =1$ and $\lambda = 0.1$ 
as a function of rescaled time $t$, where the Heisenberg time is at $t= 2\pi$. 
Numerical simulations (red points) are compared to the linear response theory 
(green solid line) and its exponentiation (blue dashed line). Panel (a) shows 
the average diagonal element $\la\varrho_\rmc^{11}(t)\ra$ and panel (b) the 
purity of $\la\varrho_\rmc(t)\ra$.}
\label{S:fig2a}\end{figure}

\paragraph{Eigenstate of \boldmath $\sigma_z$}
In Fig.~\ref{S:fig2a}, we show simulations for the reduced dynamics of the 
qubit, where the initial state is chosen as $|1\ra\la 1|$, the eigenstate of
$\sigma_z$ corresponding to the eigenvalue $+1$. The results suggest that the
evolving qubit state is of the form
\begin{equation}
\la\varrho_\rmc(t)\ra = \begin{pmatrix} r(t) & 0\\ 0 & 1-r(t)\end{pmatrix} \; ,
\end{equation}
which corresponds to a trajectory restricted to the $z$-axis in the Bloch 
sphere. For this reason, we only consider the diagonal element 
$\la\varrho_\rmc^{11}(t)\ra = r(t)$ in panel (a) of Fig.~\ref{S:fig2a}. The 
residual fluctuations of the non-diagonal elements are very small and well 
within the expected statistical error. For the purity of the average qubit 
state, we therefore find:
\begin{equation}
P(t)= r(t)^2 + [1 - r(t)]^2 = r^2 + 1 + r^2 - 2r = 1 - 2r(1-r) \; .
\end{equation}
Even though $\Delta$ and $\mu$ have been chosen as in the previous cases, the
qubit state does clearly not tend to the maximally mixed state as
it seemed to be the case before. For large times, we rather end up with a 
finite polarization: $\lim_{t\to\infty} 2r(t) - 1 \approx 55\%$.

In the present case, the linear response approximation (green solid lines) 
provides a rather unsatisfying description. The agreement with the numerical
simulation is restricted to very small times, and the exponentiation (blue
dashed lines) does not really improve the situation. For the exponentiation,
we apply the replacement
\begin{equation}
1- 2\mu^2\, C_z(t) \to b + (1-b)\; \exp[-2 \mu^2/(1-b)\, C_z(t)] \; ,
\label{S:explrsigz}\end{equation}
to Eq.~(\ref{S:tirhoz}), where the value of $b$ is fitted to the numerical 
simulation.

\subsection{\label{SS} Effect of spectral correlations}

\begin{figure}
\includegraphics[width=0.5\textwidth]{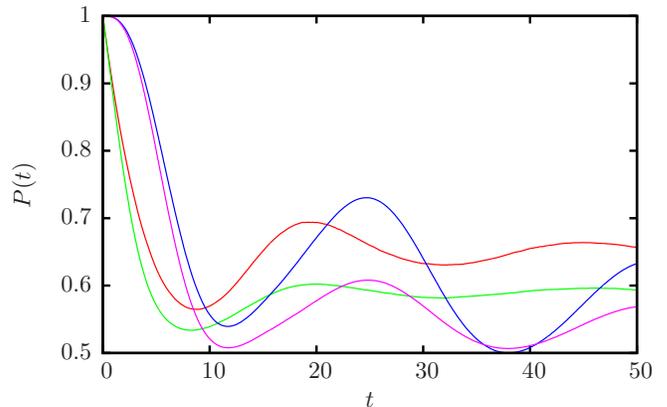}
\caption{(Color online) The purity of the average reduced qubit state, as a function of 
rescaled time $t$, for spectra with GUE correlations (red and blue lines) and 
without correlations (green and purple lines). For coupling strength and level 
splitting, we choose $\mu =0.1$ and $\Delta = 0.25$. As initial condition, we 
consider an eigenstate of $\sigma_z$ (red and green lines) and of $\sigma_x$ 
(blue and purple lines).}
\label{f:speccorrs}\end{figure}

We saw from the linear response calculation, that the dynamics of the qubit 
is affected by the type of correlations present in the spectrum of the 
Hamiltonian $h_\rme$ describing the dynamics in the environment, namely via 
the two-point form factor introduced in Eq.~(\ref{correlfun}). Quite generally,
the effect is such that spectral correlations with positive two-point function,
typical for quantum chaotic systems, lead to slower decoherence or fidelity 
decay than the spectral correlations of typical integrable systems, where the
two-point function tends to be zero~\cite{BerTab77,Haake2001}. For RMT
models, this apparently counterintuitive result has been observed 
already in Ref.~\cite{GPS04}. For dynamical systems it has been observed even 
earlier~\cite{Pro02,ProZni02} with more dramatic effects and a somewhat 
different semi-classical explanation.  

In our model which is based on unitarily invariant ensembles, only spectral 
correlations are relevant, so that we expect a possible effect to be largest
at times of the order of the Heisenberg time. This implies that the coupling
strength should be such that the decoherence time is of that size (cross-over
regime). We therefore fix the coupling strength at $\mu = 0.1$. In order to
study the effect of interest, we replaced in the diagonal Hamiltonian $h_\rme$
the unfolded GUE spectrum by independent random numbers, with the same uniform 
level density. Similar RMT ensembles with uncorrelated levels have 
been introduced in the context of statistical scattering~\cite{GDMRS97,G99}.

Fig.~\ref{f:speccorrs} shows the purity $P(t)$ as a function of time, for 
$\Delta = 0.25$ the initial state being an eigenstate of $\sigma_z$ (red
and green lines) and $\sigma_x$ (blue and purple lines). In both cases, the 
replacement of the GUE eigenvalues with uncorrelated random levels results in a 
very clear additional loss of purity. The effect is clearest at times of the 
order of $4$ times the Heisenberg time ($t\approx 25$), which is reasonable in 
view of the fact that then the resonance condition $\Delta t \approx 2\pi$ is 
fulfilled. Note that the central system can couple particularly efficiently to 
the environment, when the level splitting in the central system is in resonance 
with the transition between two levels in the environment. Comparing the level 
spacing distributions for GUE spectra (Wigner surmise) and uncorrelated spectra 
(Poisson statistics)~\cite{BerTab77,Meh2004}, we see that the probability to 
find levels with a distance $\Delta = 0.25$ is much more likely in the latter 
case. By consequence uncorrelated spectra lead to more pronounced decoherence. 

We also performed simulations with $\Delta = 1$. However, in this case the 
difference in decoherence between GUE and uncorrelated spectra was much 
smaller. This confirms our explanation, since the probability to find levels
with distances close to one, is very similar whether the spectrum shows level
repulsion or not.

\subsection{\label{SL} Accuracy of the linear response approximation}

In Ref.~\cite{GPS04} it was shown that the exponentiation of the linear 
response result yields a uniform approximation, providing a very accurate
description of the fidelity decay over the whole process. In this section, we
study the accuracy of the linear response expression for the reduced dynamics 
of the qubit in the RMT model, as described in Sec.~\ref{PX}. The 
present situation is more complicated than in Ref.~\cite{GPS04}, since there 
are now two independent parameters which determine the dynamics, the coupling 
strength $\mu$ and the level splitting $\Delta$. In addition, the object of 
interest is a two by two density matrix, not just a single function. In spite 
of these complications, we have seen that the ELR may be surprisingly accurate, 
as for instance in the case of Fig.~\ref{S:fig1a} and Fig.~\ref{S:fig3-1}.

For definiteness, we restrict ourselves to the case of eigenstates of 
$\sigma_y$ as initial states. This has the advantage that the qubit state
always tends to the maximally mixed state in the limit of large times. Thus,
it is not necessary to introduce an additional fit parameter as in
Eq.~(\ref{S:explrsigz}). Note that in Sec.~\ref{SE}, following below, we show
that eigenstates of $\sigma_x$ as initial states in general lead to stationary
states which are not maximally mixed. 

In order to quantify the accuracy of the ELR we need to quantify
the similarity between two states as they evolve in time, the first being
obtained from  numerical simulation $\varrho^{\rm ns}_\rmc(t)$ and the 
second from the ELR, $\varrho^{\rm ELR}_\rmc(t)$. For that purpose, we use the 
trace distance~\cite{Fuchs97,GiLaNie05}, defined as
\begin{equation}
D[\varrho^{\rm ns}_\rmc(t),\varrho^{\rm ELR}_\rmc(t)]=
{\rm tr}|\varrho^{\rm ns}_\rmc(t)-\varrho^{\rm ELR}_\rmc(t)| \; ,
\label{def-trace-dist}\end{equation}
where the absolute value $|A|$ of an operator $A$ stands for 
$|A|= \sqrt{A^\dagger A}$. For a Hermitian matrix $A$, the trace of $|A|$ can 
be calculated conveniently as the sum of absolute values of the eigenvalues of 
$A$. For a two level system, it is straight forward to verify that the trace
distance between two quantum states is proportional to their Euclidean distance
in the Bloch sphere representation. Finally, in order to obtain a single number
characterizing the similarity between two evolutions, we choose the maximum
value of the distance in Eq.~(\ref{def-trace-dist}), taken over the full
time range. This quantity is considered as a function of the parameters
$\Delta$ and $\mu$,
\begin{equation}
\mathcal{D}_{\rm max}(\Delta,\mu)=
\max_t D[\varrho^{\rm ns}_\rmc(t),\varrho^{\rm ELR}_\rmc(t)] \; .
\label{max-trace-dist}\end{equation}

\begin{figure}
\includegraphics[width= 0.5\textwidth]{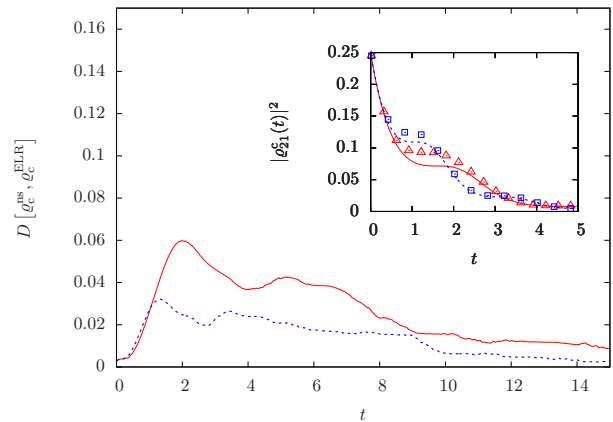}
\caption{(Color online) Trace distance between $\varrho^{\rm ns}_\rmc(t)$ and 
$\varrho^{\rm ELR}_\rmc(t)$, for the initial state being the symmetric 
eigenstate of $\sigma_y$ and $\mu=0.25$. Red solid line corresponds to 
$\Delta=1$ and the blue dotted line to $\Delta=1.5$. The inset shows the 
modulus squared of the average non-diagonal matrix element of the qubit state. 
There, the red triangles (blue squares) and the red solid line (blue dotted 
line) correspond to the numerical simulation and the ELR, respectively.}
\label{fig:tra-dist}\end{figure}

In Fig.~\ref{fig:tra-dist} we illustrate the application of our distance 
measure to two test cases, with $\Delta=1$ (red solid line) and $\Delta=1.5$ 
(blue dotted line), while $\mu= 0.25$ in both cases. This is the only figure, 
where we increase the dimension of the environment to $N_\rme=300$, in order to 
suppress residual finite size effects at very small times. It can be seen
that the distance between the ELR approximation and the numerical
simulation is varying strongly over time, with a general tendency to diminish
towards larger times, where the qubit state approaches the completely mixed
state.  The quantity $\mathcal{D}_{\rm max}$ defined in 
Eq.~(\ref{max-trace-dist}) will select the global maximum value of each curve.
The inset of Fig.~\ref{fig:tra-dist} shows the modulus squared of the 
non-diagonal element of the average qubit state. There, we compare ELR 
approximations and numerical simulations on an absolute scale.

\begin{figure}
\includegraphics[width= 0.5\textwidth]{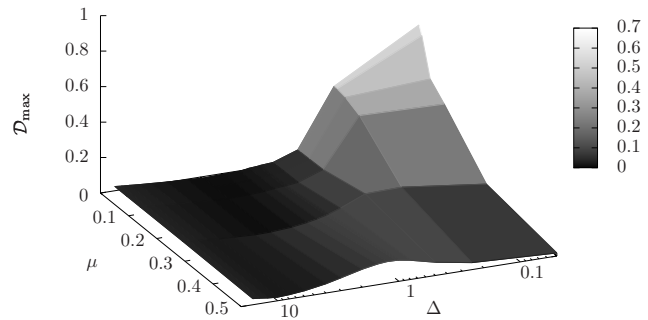}
\caption{Maximum trace distance between ELR approximation and numerical 
simulations as a function of the coupling strength $\mu$ and the energy 
splitting $\Delta$, for the initial state being an eigenstate of $\sigma_y$.}
\label{fig:max-tra-dist}\end{figure}

In Fig.~\ref{fig:max-tra-dist}, we quantify the accuracy of the ELR result as a 
uniform approximation over the whole decoherence process. For that purpose,
we present a surface plot of $\mathcal{D}_{\rm max}(\Delta,\mu)$, based on 102 
different points in the parameter space of $\mu$ and $\Delta$. We can roughly 
distinguish two regions, one, where the ELR approximation works reasonably 
well, and one where it fails. The latter region can be located in the rectangle 
$0 < \mu < 0.3$ and $0 < \Delta < 0.8$, while everywhere else 
$D_{\rm max}$ is at least smaller than $0.2$. Going away from that rectangular
region, either by increasing $\Delta$ (keeping $\mu$ fixed), or by increasing 
$\mu$ (keeping $\Delta$ fixed), always results in the reduction of 
$D_{\rm max}$ to very small values. The slowest decay of the error can be 
observed for increasing $\mu$ at $\Delta \approx 1$. However, also there 
$D_{\rm max}$ becomes ever smaller when $\mu$ is further increased as we 
checked with additional simulations. 

We may explain the success of the ELR approximation in the different limits as
follows: For fixed $\Delta$, when $\mu$ is increased, we arrive at the Fermi
golden rule regime, where the ELR approximation becomes exact. For fixed $\mu$,
where $\Delta$ is increased, the double commutator in Eq.~(\ref{eq-principal}) 
has the function $\cos(\Delta s')$ as a common prefactor, see 
Eq.~(\ref{S:doublecomcos}), such that the fast oscillations tend to suppress the 
$s'$-integral except for that part of the correlation function $c(s-s')$ which 
contains de $\delta$-function. In this way we arrive again at the Fermi golden 
rule result. 
The fact that the ELR approximation fails in the limit of small $\Delta$ (for
$\mu \lesssim 0.3$) and small $\mu$ (for $\Delta \lesssim 1$) suggests, that 
the ELR approximation is not consistent with a treatment on the basis of 
time-independent perturbation theory.

\subsection{\label{SE} Symmetry breaking stationary state}

In this section, we investigate the stationary states of our model, 
$\la\varrho_\rmc(t)\ra$ in the limit $t\to\infty$. 
For some choices of the parameters $\Delta$ and $\mu$, the system ends up in
different stationary states, depending on the initial state. In other words,
the system may remain in a given state (if so, this state would be called 
stationary) or not depending on the history of the evolution. This is very
clearly a memory effect and a signature for non-Markovianity~\cite{BreuPet02}. 
For the following discussion, the representation of the qubit states 
$\la\varrho_\rmc(t)\ra$ as points in the Bloch sphere will again prove useful.

Since the system Hamiltonian is proportional to $\sigma_z$, any qubit state at
a finite distance of the $z$-axis will experience a torque, which can be
computed from the Ehrenfest theorem. In the absence of any counteracting force, 
such a state will start to rotate around the $z$-axis. Therefore, one would 
expect that any stationary state must be located on the $z$-axis itself.
Our numerical simulations show that this is not always the case. There are
situations, where the coupling to the RMT environment breaks the rotational 
symmetry of the stationary state, which then settles on the $x$-axis at a 
finite distance from the origin.

For any state, 
starting out in the $yz$-plane of the Bloch sphere, the average qubit state 
$\la\varrho_\rmc(t)\ra$ will finally end up on the $z$-axis. As one can see 
from Fig.~\ref{S:fig2a}, starting out from an eigenstate of $\sigma_z$ (this 
corresponds to either the South ($z=-1$) or the North pole ($z=1$) on the Bloch 
sphere, the evolution of the qubit state is always restricted to the $z$-axis,
and the stationary state will typically be found at a finite distance from the 
origin, which represents the uniformly mixed state. By contrast, starting out 
on the $y$-axis, the evolution of $\la\varrho_\rmc(t)\ra$ is restricted to the 
$xy$-plane and the stationary state is precisely the uniformly mixed state, cf. 
Fig.~\ref{S:fig3-1}. Because of the linearity of the mapping 
$\varrho_\rmc \to \la\varrho_\rmc(t)\ra$, any state in the $yz$-plane will be
mapped on the $z$-axis as $t$ goes to infinity.

\begin{figure}
\includegraphics[width= 0.75\textwidth]{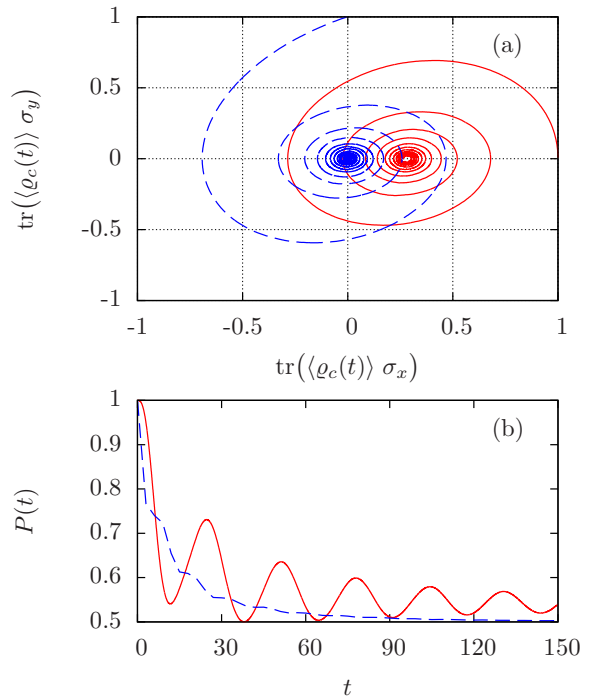}
\caption{(Color online) Decoherence process for $\Delta= 0.25$ and 
$\mu= 0.1$ for the initial state being an eigenstate of $\sigma_x$ (red solid 
curves) and of $\sigma_y$ (blue dashed curves). Panel (a): The trajectory of 
the average qubit state $\la\varrho_\rmc(t)\ra$ remains in the $xy$ plane of 
the Bloch sphere.  Panel (b): Purity as a function of time for the same cases.}
\label{fig:move-equi-st}\end{figure}

It came as a surprise that in the case of a $\sigma_x$ eigenstate as initial
state, it is possible that $\la\varrho_\rmc(t)\ra$ converges to a state on the
$x$-axis at a finite distance from the origin.
This distance depends on the parameters $\Delta$ and $\mu$ and might be very 
small, as for instance in Fig.~\ref{S:fig1a}. However, choosing these 
parameters appropriately, the distance can be quite notable, as in 
Fig.~\ref{fig:move-equi-st}, panel (a). In Fig.~\ref{fig:move-equi-st} we 
analyse the trajectory of the qubit state $\la\varrho_\rmc(t)\ra$ in the Bloch
sphere, for $\Delta= 0.25$ and $\mu= 0.1$, for
an eigenstate of $\sigma_x$ (red solid line) and an eigenstate of $\sigma_y$
(blue dashed line) as initial state. Both trajectories are restricted to the 
$xy$-plane ($z=0$), however, while the blue dashed line converges to the origin 
of the plane, i.e. the completely mixed state, the red solid line clearly 
converges to a point on the $x$-axis approximately 0.3 units away from the 
origin. In panel (b), we show the purity of the corresponding two states. While 
the blue dashed line decreases with very small undulations towards the limit 
value $P(t) \to 1/2$, the red solid line shows strong oscillations, and seems 
to setlle on a value around $P(t) \to 0.55$. Note however, that the convergence 
is extremely slow.

As explained above, the only possibility for such a state to be a stationary 
state consists in the cancellation of the torque applied by the central system 
Hamiltonian due to the coupling term. Note however that $\sigma_x$
cannot produce this compensation directly, as can be deduced by applying 
the Ehrenfest theorem, again. Instead, the compensating force must be exerted 
via the environment part of the state during the unitary dynamics in the full 
system (qubit plus environment). 
This is somewhat unexpected, because the environment part of the coupling 
operator is on average unitarily invariant, yet the individual terms in the 
ensemble are not.

Finally, note that Fig.~\ref{fig:move-equi-st} illustrates again the 
non-Markovianity of the evolution of the qubit. As we can see, the two 
different initial conditions lead to trajectories (the blue and the red line in 
upper panel) which intersect very often, but finally approach different 
equilibrium states. It is not shown here, but it is not difficult to find cases 
where the intersection happens exactly at the same time. In such a case, the 
evolution continues in different directions, even though the state is the same. 
Thus the continuation also depends on where the trajectory did come from -- a 
clear manifestation of a memory effect.

\section{\label{C} Conclusions}

Previous random matrix models for the environment and its coupling to the 
central system, e.g.~\cite{PGS07,GPKS08}, assumed the lack of any knowledge
about the coupling between central system and environments, thus yielding a 
generic result, equivalent to an average over different separable 
couplings. In this paper we identified different types of such couplings 
and derived an analytical description in the linear response approximation.

Concentrating on one particularly interesting case, we found
a variety of rather unexpected features with a central system as 
simple as a single qubit. We developed an analytical description based on 
linear response theory, and investigated its accuracy. We described several
effects in greater detail, such as purity oscillations, non-Markovian 
dynamics, and symmetry breaking stationary states. Some of these are clearly 
not of perturbative nature as we find them in numerics but not in our linear 
response solutions. 
 
In future work we might consider different symmetry classes for the random 
matrix environment, as well as different couplings, e.g. similar to the 
Jaynes-Cummings coupling. Most importantly, we will try to improve the 
analytical description, e.g. incorporating time-independent perturbation
theory. We also plan to study in more detail the degree of non-Markovianity, 
and the possibility to use time-local master equations for an accurate 
description of the dynamics of this open system.

\begin{acknowledgments}
We thank C. Pineda and F. Toscano for enlightening discussions, and we 
acknowledge the hospitality of the Centro Internacional de Ciencias, UNAM, 
where many of these discussions took place. We also acknowledge financial 
support from CONACyT through the grants 129309 and 154586 as well as 
UNAM/DGAPA/PAPIIT IG 101113.
\end{acknowledgments}

\begin{widetext}
\appendix*
\section{\label{apE} Evaluation of integrals}
In the present paper, we restrict the linear response calculations to the GUE 
case, where the basic correlation function reads:
\begin{equation}
c(t) = 1 + \delta(\tau) - b_2(\tau) \; , \qquad 
b_2(\tau)= \begin{cases} 
           1- \tau &: \tau < 1\\ 0 &: \tau > 1
            \end{cases} \; , 
\end{equation}
where $\tau=t/(2\pi)$ and we have assumed that $t>0$.
In this section we evaluate integrals which describe the evolution of the
matrix elements of the density matrix of a qubit in contact with a random
matrix environment in the interaction picture. The results are required in 
Sec.~\ref{S}.

\subsection{\label{apED} Dephasing coupling \boldmath $v_\rmc = \sigma_z$}

The evolution of the relevant quantity for this case, $\tilde\varrho_{21}(t)$,
is given by the decay of a fidelity 
amplitude -- see Eq.~(\ref{S:defCfid}). We therefore denote the integral to be 
evaluated by $C_{\rm fid}(t)$. By changing the integration variable from 
$s'$ to $u= s - s'$, we obtain:
\begin{align}
C_{\rm fid}(t) &= \int_0^t\rmd s\int_0^s\rmd s'\; c(s-s')
 = \int_0^t\rmd u\; c(u)\; (t-u)
 = \frac{t^2}{2} + \pi\, t - \int_0^{\min(t,2\pi)}\rmd u\; 
   \left( 1 - \frac{u}{2\pi}\right) (t-u) \notag\\
&= \frac{t^2}{2} + \pi\, t - \begin{cases} t^2/2 - t^3/(12\pi) &: t < 2\pi\\
      \pi\, t - 2\pi^2/3 &: t > 2\pi\end{cases}
 = \begin{cases} \pi\, t + t^3/(12\pi) &: t < 2\pi\\  
      t^2/2 + 2\pi^2/3 &: t > 2\pi \end{cases} \; .
\label{apE:resCfid}\end{align}

\subsection{Coupling operator \boldmath $v_\rmc = \sigma_x$}

In this section, we evaluate the integrals required for the description of
the non-commutative case, where the coupling operator is 
$\sigma_x\otimes V_\rme$, which does not commute with the Hamiltonian of the
central system.

\subsubsection{The initial state of the central system being an eigenstate of 
  \boldmath $\sigma_x$}

In order to calculate $C_x(t)$ in Eq.~(\ref{S:defCx}), we first calculate
\begin{align}
b_x(u) &= \int_u^t\rmd s\; \rme^{-\rmi\Delta\, s}\; \rmi\; \sin\, \Delta(s-u)
 = \frac{1}{4\Delta}\, \Big [\, \big (\, \rmi + 2\Delta\, (t-u)\, \big )\; 
   \rme^{-\rmi\Delta u} - \rmi\; \rme^{\rmi\Delta\, (u - 2t)} \, \Big ] 
\end{align}
and then divide the correlation function $c(u)$ in two parts, so that
\begin{equation}
C_x(t)= \bar C_x(t) - B_x(t) \; ,
\end{equation}
where
\begin{align}
\bar C_x(t) &= \int_0^t\rmd u\; \big \{\, 1 + \delta[u/(2\pi)]\, \big \}\;
     b_x(u) 
 = \int_0^t\rmd u\; b_x(u) + \pi\; b_x(0)\notag\\
B_x(t) &= \int_0^t\rmd u\; b_2(u)\; b_x(u) 
 = \int_0^{{\rm min}(t,2\pi)}\rmd u\; 
   \Big (\, 1 - \frac{u}{2\pi}\, \Big )\; b_x(u) \; .
\end{align}
This yields:
\begin{equation}
\bar C_x(t) = \frac{1}{4\Delta^2}\Big [\, 
    4\; \big (\, 1 - \rme^{-\rmi\Delta\, t}\, \big )
  - (1-\rmi\pi\Delta)\;
 \big (\, 1 - \rme^{-2\rmi\Delta\, t}\, \big ) 
  - 2\Delta\, t\; (\rmi - \pi\Delta)\, \Big ],
\label{apE:resCx0}
\end{equation}
the integral over the two-point form factor yields:
\begin{multline}
B_x(t) = \frac{1}{4\pi\Delta^3}\Bigg [\, 
(1 - 2\pi\rmi\Delta)\; \Delta\, t
  + 3\pi\, \Delta + \frac{5\, \rmi}{2}
+ \left(\pi\Delta - \frac{\rmi}{2}\right)\; 
  \rme^{-2\rmi\Delta\, t} \\
+ \left. \begin{cases}
 \big [\, 2\Delta\, (t-2\pi) - 2\rmi\, \big ]\; \rme^{-\rmi\Delta t} &: 
    t < 2\pi\\[0.3cm]
 \left [\, \Delta\, (2\pi -t) - \dfrac{5\rmi}{2}\, \right]\; \rme^{-2\pi\rmi\, \Delta}
   + \dfrac{\rmi}{2}\; \rme^{2\rmi\Delta (\pi-t)}  &: t > 2\pi
   \end{cases} \quad \right] \; .
\label{apE:resBx}
\end{multline}

\subsubsection{The initial state of the central system being an eigenstate of 
\boldmath $\sigma_y$ }

Similar to the procedure in the case of the $\sigma_x$ coupling, we first
calculate
\begin{align}
b_y(u) = \int_u^t\rmd s\; \rme^{-\rmi\Delta\, s}\; \cos\, \Delta (s-u)
 = \frac{1}{4\Delta}\left\{\, [2\Delta\, (t-u) - \rmi]\; \rme^{-\rmi\Delta\, u}
      + \rmi\; \rme^{\rmi\Delta\, (u-2t)}\, \right\} \; .
\end{align}
Dividing the correlation function $c(u)$ in two parts,
\begin{equation}
C_y(t)= \bar C_y(t) - B_y(t) \; , 
\end{equation}
where
\begin{align}
\bar C_y(t)& = \int_0^t\rmd u\; \big \{\, 1 - \delta[u/(2\pi)]\, \big \}\;
     b_y(u)
 = \int_0^t\rmd u\; b_y(u) + \pi\; b_y(0)\notag\\
\intertext{and}
B_y(t) &= \int_0^t\rmd u\; b_2(u)\; b_y(u)
 = \int_0^{{\rm min}(t,2\pi)}\rmd u\; 
   \Big (\, 1 - \frac{u}{2\pi}\, \Big )\; b_y(u) \; .
\end{align}
The evaluation of these integrals yield:
\begin{equation}
\bar C_y(t)= \frac{1}{4\Delta^2}\Big [\, 2\Delta\, t\; (\pi\Delta - \rmi)
  + (1- \rmi\pi\Delta)\; \big (\, 1 - \rme^{-2\rmi\Delta\, t}\, \big )\, \Big ]
\; ,
\label{apE:resCy0}\end{equation}
and
\begin{equation}
B_y(t) = \frac{1}{4\pi\Delta^3}\left[\, (1-2\pi\rmi\, \Delta)\, \Delta\, t
  + \pi\Delta + \dfrac{3\rmi}{2} + \left(\dfrac{\rmi}{2} - \pi\Delta
\right)\; 
\rme^{-2\rmi\Delta\, t}
\begin{cases} 
   2\rmi\; \rme^{-\rmi\Delta\, t} &: t < 2\pi\\[10.0pt]
   \dfrac{\rmi}{2}\; \rme^{2\rmi\Delta (\pi -t)} + \left[
\dfrac{3\rmi}{2} - (2\pi -t)\, \Delta\right]\;
   \rme^{-2\pi\rmi\Delta} &: t > 2\pi\end{cases} \right] \; .
\label{apE:resBy}
\end{equation}

\subsubsection{The initial state of the central system being an eigenstate of 
  \boldmath $\sigma_z$}
In this case
\begin{align}
C_z(t)&= \int_0^t\rmd s\; \int_0^s\rmd u\; c(u)\; 
\cos(\Delta u)
= \int_0^t\rmd u\; c(u)\; \cos(\Delta u)\; (t-u) \; .
\label{apE:resC3x0}
\end{align}
Separating the integral in the same way as above, we find
\begin{equation}
C_z(t)= \bar C_z(t) - B_z(t) \; , \quad
\text{where}\quad
\bar C_z(t)= \frac{1 - \cos\, \Delta t}{\Delta^2} + \pi\, t
\end{equation}
and
\begin{align}
B_z(t) &= \int_0^{{\rm min}(t,2\pi)}\rmd u\;  
   \left( 1 - \frac{u}{2\pi}\right)\; \cos(\Delta u)\; (t-u), \notag\\
\end{align}
evaluating the above integral we obtain finally:
\begin{equation}
B_z(t) = \frac{1}{\Delta^2}\left[\, 1 + \frac{t}{2\pi} -
\begin{cases} 
 \frac{ \sin\, \Delta t}{\pi\Delta} + \left(1 - \dfrac{t}{2\pi}\right)\; \cos\, \Delta t
      &: t < 2\pi\\[10.0pt]
\frac{\sin\, 2\pi\Delta}{\pi\Delta}-
\left(1 - \dfrac{t}{2\pi}\right)\; \cos\, 2\pi\Delta
      &: t > 2\pi \end{cases} 
\right] \; .
\label{apE:resB3x}
\end{equation}
\end{widetext}

\bibliography{/home/gorin/Bib/JabRef}

\begin{thebibliography}{10}%
\makeatletter
\providecommand \@ifxundefined [1]{%
 \ifx #1\undefined \expandafter \@firstoftwo
 \else \expandafter \@secondoftwo
\fi
}%
\providecommand \@ifnum [1]{%
 \ifnum #1\expandafter \@firstoftwo
 \else \expandafter \@secondoftwo
\fi
}%
\providecommand \enquote [1]{``#1''}%
\providecommand \bibnamefont  [1]{#1}%
\providecommand \bibfnamefont [1]{#1}%
\providecommand \citenamefont [1]{#1}%
\providecommand\href[0]{\@sanitize\@href}%
\providecommand\@href[1]{\endgroup\@@startlink{#1}\endgroup\@@href}%
\providecommand\@@href[1]{#1\@@endlink}%
\providecommand \@sanitize [0]{\begingroup\catcode`\&12\catcode`\#12\relax}%
\@ifxundefined \pdfoutput {\@firstoftwo}{%
 \@ifnum{\z@=\pdfoutput}{\@firstoftwo}{\@secondoftwo}%
}{%
 \providecommand\@@startlink[1]{\leavevmode\special{html:<a href="#1">}}%
 \providecommand\@@endlink[0]{\special{html:</a>}}%
}{%
 \providecommand\@@startlink[1]{%
  \leavevmode
  \pdfstartlink
   attr{/Border[0 0 1 ]/H/I/C[0 1 1]}%
   user{/Subtype/Link/A<</Type/Action/S/URI/URI(#1)>>}%
  \relax
 }%
 \providecommand\@@endlink[0]{\pdfendlink}%
}%
\providecommand \url  [0]{\begingroup\@sanitize \@url }%
\providecommand \@url [1]{\endgroup\@href {#1}{\urlprefix}}%
\providecommand \urlprefix [0]{URL }%
\providecommand \Eprint[0]{\href }%
\@ifxundefined \urlstyle {%
  \providecommand \doi [1]{doi:\discretionary{}{}{}#1}%
}{%
  \providecommand \doi [0]{doi:\discretionary{}{}{}\begingroup
  \urlstyle{rm}\Url }%
}%
\providecommand \doibase [0]{http://dx.doi.org/}%
\providecommand \Doi[1]{\href{\doibase#1}}%
\providecommand \bibAnnote [3]{%
  \BibitemShut{#1}%
  \begin{quotation}\noindent
    \textsc{Key:}\ #2\\\textsc{Annotation:}\ #3%
  \end{quotation}%
}%
\providecommand \bibAnnoteFile [2]{%
  \IfFileExists{#2}{\bibAnnote {#1} {#2} {\input{#2}}}{}%
}%
\providecommand \typeout [0]{\immediate \write \m@ne }%
\providecommand \selectlanguage [0]{\@gobble}%
\providecommand \bibinfo [0]{\@secondoftwo}%
\providecommand \bibfield [0]{\@secondoftwo}%
\providecommand \translation [1]{[#1]}%
\providecommand \BibitemOpen[0]{}%
\providecommand \bibitemStop [0]{}%
\providecommand \bibitemNoStop [0]{.\EOS\space}%
\providecommand \EOS [0]{\spacefactor3000\relax}%
\providecommand \BibitemShut [1]{\csname bibitem#1\endcsname}%
\bibitem{LutWei99}%
  \BibitemOpen
  \bibfield{author}{%
  \bibinfo {author} {\bibfnamefont{E.}~\bibnamefont{Lutz}}\ and\ \bibinfo
  {author} {\bibfnamefont{H.~A.}\ \bibnamefont{Weidenm\"{u}ller}},\ }%
  \bibfield{journal}{%
  \bibinfo {journal} {Physica A}\ }%
  \textbf{\bibinfo {volume} {267}},\ \bibinfo {pages} {354} (\bibinfo {year} 
  {1999})%
  \bibAnnoteFile{NoStop}{LutWei99}%
\bibitem{GS02b}%
  \BibitemOpen
  \bibfield{author}{%
  \bibinfo {author} {\bibfnamefont{T.}~\bibnamefont{Gorin}}\ and\ \bibinfo
  {author} {\bibfnamefont{T.~H.}\ \bibnamefont{Seligman}},\ }%
  \bibfield{journal}{%
  \bibinfo {journal} {J. Opt. B: Quant. Semiclass. Opt.}\ }%
  \textbf{\bibinfo {volume} {4}},\ \bibinfo {pages} {S386} (\bibinfo {year}
  {2002})%
  \bibAnnoteFile{NoStop}{GS02b}%
\bibitem{GS03}%
  \BibitemOpen
  \bibfield{author}{%
  \bibinfo {author} {\bibfnamefont{T.}~\bibnamefont{Gorin}}\ and\ \bibinfo
  {author} {\bibfnamefont{T.~H.}\ \bibnamefont{Seligman}},\ }%
  \bibfield{journal}{%
  \bibinfo {journal} {Phys. Lett. A}\ }%
  \textbf{\bibinfo {volume} {309}},\ \bibinfo {pages} {61} (\bibinfo {year} 
  {2003})%
  \bibAnnoteFile{NoStop}{GS03}%
\bibitem{Zni11}%
  \BibitemOpen
  \bibfield{author}{%
  \bibinfo {author} 
  {\bibfnamefont{M.}~\bibnamefont{\v{Z}nidari\v{c}}}, 
  \bibinfo {author}
  {\bibfnamefont{C.}~\bibnamefont{Pineda}},\ and\ \bibinfo {author}
  {\bibfnamefont{I.}~\bibnamefont{Garc\'{\i}a-Mata}},\ }%
  \bibfield{journal}{%
  \Doi{10.1103/PhysRevLett.107.080404}{\bibinfo {journal} {Phys. Rev. Lett.}}\
  }%
  \textbf{\bibinfo {volume} {107}},\ \bibinfo {pages} {080404} (\bibinfo {year} {2011})%
  \bibAnnoteFile{NoStop}{Zni11}%
\bibitem{PGS07}%
  \BibitemOpen
  \bibfield{author}{%
  \bibinfo {author} {\bibfnamefont{C.}~\bibnamefont{Pineda}}, \bibinfo {author}
  {\bibfnamefont{T.}~\bibnamefont{Gorin}},\ and\ \bibinfo {author}
  {\bibfnamefont{T.~H.}\ \bibnamefont{Seligman}},\ }%
  \bibfield{journal}{%
  \bibinfo {journal} {New J. Phys.}\ }%
  \textbf{\bibinfo {volume} {9}},\ \bibinfo {eid} {106} (\bibinfo {year}
  {2007})%
  \bibAnnoteFile{NoStop}{PGS07}%
\bibitem{GPKS08}%
  \BibitemOpen
  \bibfield{author}{%
  \bibinfo {author} {\bibfnamefont{T.}~\bibnamefont{Gorin}}, \bibinfo {author}
  {\bibfnamefont{C.}~\bibnamefont{Pineda}}, \bibinfo {author}
  {\bibfnamefont{H.}~\bibnamefont{Kohler}},\ and\ \bibinfo {author}
  {\bibfnamefont{T.~H.}\ \bibnamefont{Seligman}},\ }%
  \bibfield{journal}{%
  \bibinfo {journal} {New J. Phys.}\ }%
  \textbf{\bibinfo {volume} {10}},\ \bibinfo {eid} {115016} (\bibinfo {year}
  {2008})%
  \bibAnnoteFile{NoStop}{GPKS08}%
\bibitem{VinZni12}%
  \BibitemOpen
  \bibfield{author}{%
  \bibinfo {author} {\bibnamefont{Vinayak}}\ and\ \bibinfo {author}
  {\bibfnamefont{M.}~\bibnamefont{\v{Z}nidari\v{c}}},\ }%
  \bibfield{journal}{%
  \bibinfo {journal} {J. Phys. A: Math. Theor.}\ }%
  \textbf{\bibinfo {volume} {45}} (\bibinfo {year} {2012})%
  \bibAnnoteFile{NoStop}{VinZni12}%
\bibitem{NoAlJe09}%
  \BibitemOpen
  \bibfield{author}{%
  \bibinfo {author} {\bibfnamefont{J.}~\bibnamefont{Novotn\'{y}}}, \bibinfo
  {author} {\bibfnamefont{G.}~\bibnamefont{Alber}},\ and\ \bibinfo {author}
  {\bibfnamefont{I.}~\bibnamefont{Jex}},\ }%
  \bibfield{journal}{%
  \bibinfo {journal} {J. Phys. A: Math. Theor.}\ }%
  \textbf{\bibinfo {volume} {42}},\ \bibinfo {pages} {282003} (\bibinfo {year}
  {2009})%
  \bibAnnoteFile{NoStop}{NoAlJe09}%
\bibitem{NoAlJe11}%
  \BibitemOpen
  \bibfield{author}{%
  \bibinfo {author} {\bibfnamefont{J.}~\bibnamefont{Novotn\'{y}}}, \bibinfo
  {author} {\bibfnamefont{G.}~\bibnamefont{Alber}},\ and\ \bibinfo {author}
  {\bibfnamefont{I.}~\bibnamefont{Jex}},\ }%
  \bibfield{journal}{%
  \bibinfo {journal} {New J. Phys.}\ }%
  \textbf{\bibinfo {volume} {13}},\ \bibinfo {pages} {053052} (\bibinfo {year}
  {2011})%
  \bibAnnoteFile{NoStop}{NoAlJe11}%
\bibitem{Bruzda2009320}%
  \BibitemOpen
  \bibfield{author}{%
  \bibinfo {author} {\bibfnamefont{W.}~\bibnamefont{Bruzda}}, \bibinfo {author}
  {\bibfnamefont{V.}~\bibnamefont{Cappellini}}, \bibinfo {author}
  {\bibfnamefont{H.-J.}\ \bibnamefont{Sommers}},\ and\ \bibinfo {author}
  {\bibfnamefont{K.}~\bibnamefont{\.{Z}yczkowski}},\ }%
  \bibfield{journal}{%
  \Doi{http://dx.doi.org/10.1016/j.physleta.2008.11.043}{\bibinfo {journal}
  {Phys. Lett. A}}\ }%
  \textbf{\bibinfo {volume} {373}},\ \bibinfo {pages} {320 } (\bibinfo {year}
  {2009})%
  \bibAnnoteFile{NoStop}{Bruzda2009320}%
\bibitem{PhysRevA.87.022111}%
  \BibitemOpen
  \bibfield{author}{%
  \bibinfo {author} {\bibfnamefont{M.}~\bibnamefont{Musz}}, \bibinfo {author}
  {\bibfnamefont{M.}~\bibnamefont{Ku\ifmmode~\acute{s}\else \'{s}\fi{}}},\ and\
  \bibinfo {author} {\bibfnamefont{K.}~\bibnamefont{\.{Z}yczkowski}},\ }%
  \bibfield{journal}{%
  \Doi{10.1103/PhysRevA.87.022111}{\bibinfo {journal} {Phys. Rev. A}}\ }%
  \textbf{\bibinfo {volume} {87}},\ \bibinfo {pages} {022111} (\bibinfo {year} {2013})%
  \bibAnnoteFile{NoStop}{PhysRevA.87.022111}%
\bibitem{Wang:12}%
  \BibitemOpen
  \bibfield{author}{%
  \bibinfo {author} {\bibfnamefont{J.}~\bibnamefont{Wang}},\ }%
  \bibfield{journal}{%
  \Doi{10.1364/JOSAB.29.000075}{\bibinfo {journal} {J. Opt. Soc. Am. B}}\ }%
  \textbf{\bibinfo {volume} {29}},\ \bibinfo {pages} {75} (\bibinfo {year} {2012})%
  \bibAnnoteFile{NoStop}{Wang:12}%
\bibitem{GCZ97}%
  \BibitemOpen
  \bibfield{author}{%
  \bibinfo {author} {\bibfnamefont{S.~A.}\ \bibnamefont{Gardiner}}, \bibinfo
  {author} {\bibfnamefont{J.~I.}\ \bibnamefont{Cirac}},\ and\ \bibinfo {author}
  {\bibfnamefont{P.}~\bibnamefont{Zoller}},\ }%
  \bibfield{journal}{%
  \bibinfo {journal} {Phys. Rev. Lett.}\ }%
  \textbf{\bibinfo {volume} {79}},\ \bibinfo {pages} {4790} (\bibinfo {year} {1997})%
  \bibAnnoteFile{NoStop}{GCZ97}%
\bibitem{GPSS04}%
  \BibitemOpen
  \bibfield{author}{%
  \bibinfo {author} {\bibfnamefont{T.}~\bibnamefont{Gorin}}, \bibinfo {author}
  {\bibfnamefont{T.}~\bibnamefont{Prosen}}, \bibinfo {author}
  {\bibfnamefont{T.~H.}\ \bibnamefont{Seligman}},\ and\ \bibinfo {author}
  {\bibfnamefont{W.~T.}\ \bibnamefont{Strunz}},\ }%
  \bibfield{journal}{%
  \bibinfo {journal} {Phys. Rev. A}\ }%
  \textbf{\bibinfo {volume} {70}},\ \bibinfo {pages} {042105} (\bibinfo {year} {2004})%
  \bibAnnoteFile{NoStop}{GPSS04}%
\bibitem{KimMan76}%
  \BibitemOpen
  \bibfield{author}{%
  \bibinfo {author} {\bibfnamefont{H.~J.}\ \bibnamefont{Kimble}}\ and\ \bibinfo
  {author} {\bibfnamefont{L.}~\bibnamefont{Mandel}},\ }%
  \bibfield{journal}{%
  \Doi{10.1103/PhysRevA.13.2123}{\bibinfo {journal} {Phys. Rev. A}}\ }%
  \textbf{\bibinfo {volume} {13}},\ \bibinfo {pages} {2123} (\bibinfo {year} {1976})%
  \bibAnnoteFile{NoStop}{KimMan76}%
\bibitem{jaynescummings1963}%
  \BibitemOpen
  \bibfield{author}{%
  \bibinfo {author} {\bibfnamefont{E.~T.}~\bibnamefont{Jaynes}}\ and\ \bibinfo
  {author} {\bibfnamefont{F.~W.}\ \bibnamefont{Cummings}},\ }%
  \bibfield{journal}{%
  \bibinfo {journal} {Proc. IEEE}\ }%
  \textbf{\bibinfo {volume} {51}},\ \bibinfo{pages} {89} (\bibinfo {year} 
    {1963})%
  \bibAnnoteFile{NoStop}{jaynescummings1963}%
\bibitem{Wigner1967}%
  \BibitemOpen
  \bibfield{author}{%
  \bibinfo {author} {\bibfnamefont{E.~P.}\ \bibnamefont{Wigner}},\ }%
  \bibfield{journal}{%
  \bibinfo {journal} {SIAM Rev.}\ }%
  \textbf{\bibinfo {volume} {9}},\ \bibinfo {pages} {1} (\bibinfo {year} 
    {1967})%
  \bibAnnoteFile{NoStop}{Wigner1967}%
\bibitem{Balian}%
  \BibitemOpen
  \bibfield{author}{%
  \bibinfo {author} {\bibfnamefont{R.}\ \bibnamefont{Balian}},\ }%
  \bibfield{journal}{%
  \bibinfo {journal} {Nuovo Cimento}\ }%
  \textbf{\bibinfo {volume} {57B}},\ \bibinfo {pages} {183} (\bibinfo {year} 
    {1967})%
  \bibAnnoteFile{NoStop}{Balian}%
\bibitem{Casati}%
  \BibitemOpen
  \bibfield{author}{%
  \bibinfo {author} {\bibfnamefont{G.}~\bibnamefont{Casati}}, \bibinfo
  {author} {\bibfnamefont{F.}\ \bibnamefont{Valz-Gris}},\ and\ \bibinfo
  {author} {\bibfnamefont{I.}~\bibnamefont{Guarneri}},\ }%
  \bibfield{journal}{%
  \bibinfo {journal} {Lett. Nuovo Cimento}\ }%
  \textbf{\bibinfo {volume} {28}},\ \bibinfo {pages} {279} (\bibinfo {year} 
    {1980})%
  \bibAnnoteFile{NoStop}{Casati}%
\bibitem{BGS84}%
  \BibitemOpen
  \bibfield{author}{%
  \bibinfo {author} {\bibfnamefont{O.}~\bibnamefont{Bohigas}}, \bibinfo
  {author} {\bibfnamefont{M.~J.}\ \bibnamefont{Giannoni}},\ and\ \bibinfo
  {author} {\bibfnamefont{C.}~\bibnamefont{Schmit}},\ }%
  \bibfield{journal}{%
  \bibinfo {journal} {Phys. Rev. Lett.}\ }%
  \textbf{\bibinfo {volume} {52}},\ \bibinfo {pages} {1} (\bibinfo {year}
  {1984})%
  \bibAnnoteFile{NoStop}{BGS84}%
\bibitem{Lem12}%
  \BibitemOpen
  \bibfield{author}{%
  \bibinfo {author} {\bibfnamefont{G.~B.}\ \bibnamefont{Lemos}}, \bibinfo
  {author} {\bibfnamefont{R.~M.}\ \bibnamefont{Gomes}}, \bibinfo {author}
  {\bibfnamefont{S.~P.}\ \bibnamefont{Walborn}}, \bibinfo {author}
  {\bibfnamefont{P.~H.}\ \bibnamefont{Souto~Ribeiro}},\ and\ \bibinfo {author}
  {\bibfnamefont{F.}~\bibnamefont{Toscano}},\ }%
  \bibfield{journal}{%
  \bibinfo {journal} {Nat. Commun.}\ }%
  \textbf{\bibinfo {volume} {3}},\ \bibinfo {pages} {1211} (\bibinfo {year} {2012})%
  \bibAnnoteFile{NoStop}{Lem12}%
\bibitem{BerTab77}%
  \BibitemOpen
  \bibfield{author}{%
  \bibinfo {author} {\bibfnamefont{M.~V.}\ \bibnamefont{Berry}}\ and\ \bibinfo
  {author} {\bibfnamefont{M.}~\bibnamefont{Tabor}},\ }%
  \bibfield{journal}{%
  \bibinfo {journal} {Proc. R. Soc. Lond. A}\ }%
  \textbf{\bibinfo {volume} {356}},\ \bibinfo {pages} {375} (\bibinfo {year}
  {1977})%
  \bibAnnoteFile{NoStop}{BerTab77}%
\bibitem{BenZyc06}%
  \BibitemOpen
  \bibfield{author}{%
  \bibinfo {author} {\bibfnamefont{I.}~\bibnamefont{Bengtsson}}\ and\ \bibinfo
  {author} {\bibfnamefont{K.}~\bibnamefont{\.{Z}yczkowski}},\ }%
  \emph{\bibinfo {title} {Geometry of Quantum States: An Introduction to
  Quantum Entanglement}}\ (\bibinfo {publisher} {Cambridge University Press},%
  \ \bibinfo {address} {Cambridge},\ \bibinfo {year} {2006})%
  \bibAnnoteFile{NoStop}{BenZyc06}%
\bibitem{BrLaPi09}%
  \BibitemOpen
  \bibfield{author}{%
  \bibinfo {author} {\bibfnamefont{H.-P.}\ \bibnamefont{Breuer}}, \bibinfo
  {author} {\bibfnamefont{E.-M.}\ \bibnamefont{Laine}},\ and\ \bibinfo {author}
  {\bibfnamefont{J.}~\bibnamefont{Piilo}},\ }%
  \bibfield{journal}{%
  \Doi{10.1103/PhysRevLett.103.210401}{\bibinfo {journal} {Phys. Rev. Lett.}}\
  }%
  \textbf{\bibinfo {volume} {103}},\ \bibinfo {pages} {210401} (\bibinfo {year} {2009})%
  \bibAnnoteFile{NoStop}{BrLaPi09}%
\bibitem{RiHuPl10}%
  \BibitemOpen
  \bibfield{author}{%
  \bibinfo {author} {\bibfnamefont{A.}~\bibnamefont{Rivas}}, \bibinfo {author}
  {\bibfnamefont{S.~F.}\ \bibnamefont{Huelga}},\ and\ \bibinfo {author}
  {\bibfnamefont{M.~B.}\ \bibnamefont{Plenio}},\ }%
  \bibfield{journal}{%
  \Doi{10.1103/PhysRevLett.105.050403}{\bibinfo {journal} {Phys. Rev. Lett.}}\
  }%
  \textbf{\bibinfo {volume} {105}},\ \bibinfo {pages} {050403} (\bibinfo {year} {2010})%
  \bibAnnoteFile{NoStop}{RiHuPl10}%
\bibitem{Meh2004}%
  \BibitemOpen
  \bibfield{author}{%
  \bibinfo {author} {\bibfnamefont{M.~L.}\ \bibnamefont{Mehta}},\ }%
  \emph{\bibinfo {title} {Random Matrices and the Statistical Theory of Energy
  Levels, 3rd Ed.}}\ (\bibinfo {publisher} {Academic Press},\ \bibinfo
  {address} {New York},\ \bibinfo {year} {2004})%
  \bibAnnoteFile{NoStop}{Meh2004}%
\bibitem{GPS04}%
  \BibitemOpen
  \bibfield{author}{%
  \bibinfo {author} {\bibfnamefont{T.}~\bibnamefont{Gorin}}, \bibinfo {author}
  {\bibfnamefont{T.}~\bibnamefont{Prosen}},\ and\ \bibinfo {author}
  {\bibfnamefont{T.~H.}\ \bibnamefont{Seligman}},\ }%
  \bibfield{journal}{%
  \bibinfo {journal} {New J. Phys.}\ }%
  \textbf{\bibinfo {volume} {6}},\ \bibinfo {pages} {20} (\bibinfo {year} {2004})%
  \bibAnnoteFile{NoStop}{GPS04}%
\bibitem{StoSch04b}%
  \BibitemOpen
  \bibfield{author}{%
  \bibinfo {author} {\bibfnamefont{H.-J.}\ \bibnamefont{St\"{o}ckmann}}\ and\
  \bibinfo {author} {\bibfnamefont{R.}~\bibnamefont{Sch\"{a}fer}},\ }%
  \bibfield{journal}{%
  \bibinfo {journal} {New J. Phys.}\ }%
  \textbf{\bibinfo {volume} {6}},\ \bibinfo {pages} {199} (\bibinfo {year} {2004})%
  \bibAnnoteFile{NoStop}{StoSch04b}%
\bibitem{StoSch05}%
  \BibitemOpen
  \bibfield{author}{%
  \bibinfo {author} {\bibfnamefont{H.-J.}\ \bibnamefont{St\"{o}ckmann}}\ and\ 
  \bibinfo {author} {\bibfnamefont{R.}~\bibnamefont{Sch\"{a}fer}},\ }%
  \bibfield{journal}{%
  \bibinfo {journal} {Phys. Rev. Lett.}\ }%
  \textbf{\bibinfo {volume} {94}},\ \bibinfo {pages} {244101} (\bibinfo {year} {2005})%
  \bibAnnoteFile{NoStop}{StoSch05}%
\bibitem{Bro81}%
  \BibitemOpen
  \bibfield{author}{%
  \bibinfo {author} {\bibfnamefont{T.~A.}\ \bibnamefont{Brody}}, \bibinfo
  {author} {\bibfnamefont{J.}~\bibnamefont{Flores}}, \bibinfo {author}
  {\bibfnamefont{J.~B.}\ \bibnamefont{French}}, \bibinfo {author}
  {\bibfnamefont{P.~A.}\ \bibnamefont{Mello}}, \bibinfo {author}
  {\bibfnamefont{A.}~\bibnamefont{Pandey}},\ and\ \bibinfo {author}
  {\bibfnamefont{S.~S.~M.}\ \bibnamefont{Wong}},\ }%
  \bibfield{journal}{%
  \bibinfo {journal} {Rev. Mod. Phys.}\ }%
  \textbf{\bibinfo {volume} {53}},\ \bibinfo {pages} {385} (\bibinfo {year} {1981})%
  \bibAnnoteFile{NoStop}{Bro81}%
\bibitem{GPSZ06}%
  \BibitemOpen
  \bibfield{author}{%
  \bibinfo {author} {\bibfnamefont{T.}~\bibnamefont{Gorin}}, \bibinfo {author}
  {\bibfnamefont{T.}~\bibnamefont{Prosen}}, \bibinfo {author}
  {\bibfnamefont{T.~H.}\ \bibnamefont{Seligman}},\ and\ \bibinfo {author}
  {\bibfnamefont{M.}~\bibnamefont{\v{Z}nidari\v{c}}},\ }%
  \bibfield{journal}{%
  \bibinfo {journal} {Phys. Rep.}\ }%
  \textbf{\bibinfo {volume} {435}},\ \bibinfo {pages} {33} (\bibinfo {year} {2006})%
  \bibAnnoteFile{NoStop}{GPSZ06}%
\bibitem{Haake2001}%
  \BibitemOpen
  \bibfield{author}{%
  \bibinfo {author} {\bibfnamefont{F.}~\bibnamefont{Haake}},\ }%
  \emph{\bibinfo {title} {Quantum Signatures of Chaos, 2nd Ed.}}\
  (\bibinfo {publisher} {Springer},\ \bibinfo {address} {Berlin},\
  \bibinfo {year} {2001})%
  \bibAnnoteFile{NoStop}{Haake2001}%
\bibitem{Pro02}%
  \BibitemOpen
  \bibfield{author}{%
  \bibinfo {author} {\bibfnamefont{T.}~\bibnamefont{Prosen}},\ }%
  \bibfield{journal}{%
  \bibinfo {journal} {Phys. Rev. E}\ }%
  \textbf{\bibinfo {volume} {65}},\ \bibinfo {pages} {036208} (\bibinfo {year} {2002})%
  \bibAnnoteFile{NoStop}{Pro02}%
\bibitem{ProZni02}%
  \BibitemOpen
  \bibfield{author}{%
  \bibinfo {author} {\bibfnamefont{T.}~\bibnamefont{Prosen}}\ and\ \bibinfo
  {author} {\bibfnamefont{M.}~\bibnamefont{\v{Z}nidari\v{c}}},\ }%
  \bibfield{journal}{%
  \bibinfo {journal} {J. Phys. A: Math. Gen.}\ }%
  \textbf{\bibinfo {volume} {35}},\ \bibinfo {pages} {1455} (\bibinfo {year} {2002})%
  \bibAnnoteFile{NoStop}{ProZni02}%
\bibitem{GDMRS97}%
  \BibitemOpen
  \bibfield{author}{%
  \bibinfo {author} {\bibfnamefont{T.}~\bibnamefont{Gorin}}, \bibinfo {author}
  {\bibfnamefont{F.-M.}\ \bibnamefont{Dittes}}, \bibinfo {author}
  {\bibfnamefont{M.}~\bibnamefont{M\"{u}ller}}, \bibinfo {author}
  {\bibfnamefont{I.}~\bibnamefont{Rotter}}\ and\ \bibinfo {author}
  {\bibfnamefont{T.~H.}\ \bibnamefont{Seligman}},\ }%
  \bibfield{journal}{%
  \bibinfo {journal} {Phys. Rev. E}\ }%
  \textbf{\bibinfo {volume} {56}},\ \bibinfo {pages} {2481} (\bibinfo {year} {1997})%
  \bibAnnoteFile{NoStop}{GDMRS97}%
\bibitem{G99}%
  \BibitemOpen
  \bibfield{author}{%
  \bibinfo {author} {\bibfnamefont{T.}~\bibnamefont{Gorin}},\ }%
  \bibfield{journal}{%
  \bibinfo {journal} {J. Phys. A: Math. Gen.}\ }%
  \textbf{\bibinfo {volume} {32}},\ \bibinfo {pages} {2315} (\bibinfo {year} {1999})%
  \bibAnnoteFile{NoStop}{G99}%
\bibitem{Fuchs97}%
  \BibitemOpen
  \bibfield{author}{%
  \bibinfo {author} {\bibfnamefont{C.~A.}\ \bibnamefont{Fuchs}}\ and\ \bibinfo
  {author} {\bibfnamefont{J.}~\bibnamefont{van~de Graaf}},\ }%
\bibinfo {howpublished} {arXiv: quant-ph/9712042} (\bibinfo {year} {1997})%
  \bibAnnoteFile{NoStop}{Fuchs97}%
\bibitem{GiLaNie05}%
  \BibitemOpen
  \bibfield{author}{%
  \bibinfo {author} {\bibfnamefont{A.}~\bibnamefont{Gilchrist}}, \bibinfo
  {author} {\bibfnamefont{N.~K.}\ \bibnamefont{Langford}},\ and\ \bibinfo
  {author} {\bibfnamefont{M.~A.}\ \bibnamefont{Nielsen}},\ }%
  \bibfield{journal}{%
  \Doi{10.1103/PhysRevA.71.062310}{\bibinfo {journal} {Phys. Rev. A}}\ }%
  \textbf{\bibinfo {volume} {71}},\ \bibinfo {pages} {062310} (\bibinfo {year} {2005})%
  \bibAnnoteFile{NoStop}{GiLaNie05}%
\bibitem{BreuPet02}%
  \BibitemOpen
  \bibfield{author}{%
  \bibinfo {author} {\bibfnamefont{H.-P.}~\bibnamefont{Breuer}},%
  \ and\ \bibinfo {author} {\bibfnamefont{F.}~\bibnamefont{Petruccione}},\ }%
  \emph{\bibinfo {title} {The Theory of Open Quantum Systems}}\
  (\bibinfo {publisher} {Oxford University Press},%
  \ \bibinfo {address} {Oxford},\ \bibinfo {year} {2002})%
  \bibAnnoteFile{NoStop}{BreuPet02}%
\end{thebibliography}%

\end{document}